\documentclass[journal,twoside]{IEEEtran}

\usepackage{enumitem}

\usepackage{cite}
\usepackage{algorithmic}
\usepackage{graphicx}
\usepackage{textcomp}
\usepackage{color}
\usepackage{float}
\usepackage{pifont}
\usepackage{url}

\usepackage{amssymb}   

\usepackage{amscd, eufrak}
\usepackage{amsthm}    
\usepackage{amsmath}
\usepackage{amsfonts,dsfont,txfonts,pxfonts}
\usepackage{bm} 
\usepackage{mathrsfs}

\usepackage{colortbl}   
\usepackage{array,booktabs,multirow,makecell}   
\usepackage{caption}
\usepackage{subcaption} 

\usepackage{lineno}
\usepackage[all]{xy}
\usepackage{tikz}
\usepackage{verbatim}
\usepackage{wrapfig}
\usepackage[section]{placeins}

\theoremstyle{plain}

\theoremstyle{definition}

\theoremstyle{remark}

\usepackage{color}
\definecolor{subsectioncolor}{rgb}{0,0.541,0.855}

\def\BibTeX{{\rm B\kern-.05em{\sc i\kern-.025em b}\kern-.08em
    T\kern-.1667em\lower.7ex\hbox{E}\kern-.125emX}}

\begin{document}

\title{A Hybrid Framework for Blood Vessel Morphology Classification: Discrete Geometry-based Tortuosity Feature Measurement, Information Gain-based Feature Selection, and Random Forest Classification }

\author{Yu Zhong\textsuperscript{*}, Jingzhi Guo\textsuperscript{*}, Luyao Li, Zehao Wang, Zhihui Yang, Yixin Lin, Weilun Fu, and Yang Wang%
\thanks{Yu Zhong and Jingzhi Guo contributed equally to this work and should be considered co-first authors.}%
\thanks{This work is supported by the National Natural Science Foundation of China (No. 12461091) and the Scientific Research Project of North China University of Technology (No. 2023YZZKY19, No. 2024NCUTYXCX104). (Corresponding authors: Yu Zhong; Yang Wang.)}%
\thanks{Y. Zhong, J. Guo, L. Li, Z. Wang and Z. Yang are with the College of Science, North China University of Technology, Beijing 100144, P.R. China (e-mail: zhongyu@ncut.edu.cn).}%
\thanks{Y. Lin, W. Fu, and Y. Wang are with the Department of Neurosurgery, Beijing Chaoyang Hospital, Capital Medical University, Beijing 100020, P.R. China (e-mail: wangyang7839@163.com).}
}

\maketitle

\begin{abstract}
    Subjective visual grading of blood vessel tortuosity relies heavily on clinical experience, while traditional distance-based indices often fail to adequately characterize three-dimensional spatial deformation. Because abnormal internal carotid artery morphology may be clinically relevant to cerebrovascular assessment and stroke-risk evaluation, objective and reproducible quantification of vascular tortuosity is of considerable importance. To address this limitation, we propose a mathematical framework for the morphological classification of the internal carotid artery (ICA-C1) segment. The framework integrates discrete geometric feature measurement, Information Gain-based feature selection, and Random Forest classification. An initial set of 13 tortuosity features is extracted from the corresponding 379 clinical vascular centerlines using discrete geometric methods and subsequently reduced to a six-feature subset consisting of $\mathcal{TI}$, $\mathcal{AC}$, $\mathcal{TC}$, $\mathcal{AC}/\mathcal{AT}$, $\mathcal{AT}$, and $\mathcal{TT}$. The framework is evaluated in two classification tasks. For binary classification of non-severe and severe tortuosity, the RF model achieves a Macro-F1 score of 0.9206. For ternary morphological grading into straight, low-tortuosity, and high-tortuosity groups, it achieves a Macro-F1 score of 0.8626. The results indicate that elongation- and curvature-related features provide strong discriminatory information for basic screening, whereas torsion-related features contribute additional information for more detailed morphological classification. Based on the RF feature-importance values, we further define a Morphological Risk Index (MRI), which provides a direct numerical reference for vascular morphology and may facilitate more objective and consistent clinical assessment.
\end{abstract}


\section{Introduction}
\label{sec:introduction}
\begin{figure}[!t]
    \centering
    \includegraphics[width=\linewidth]{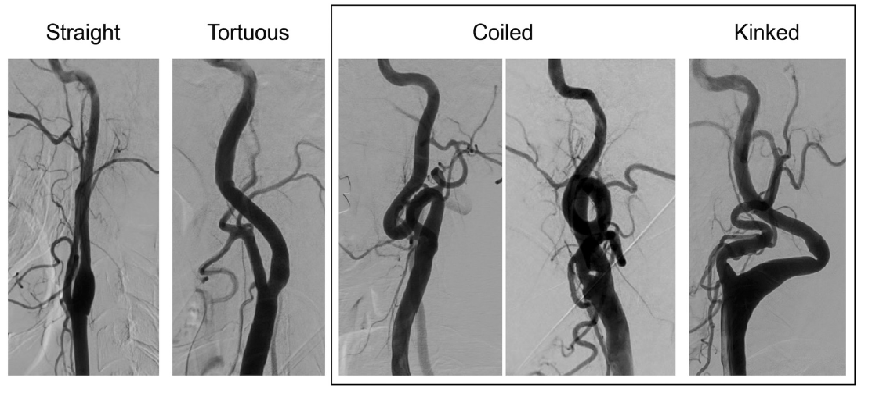} 
    \caption{Traditional clinical qualitative classification of the ICA-C1 segment based on visual inspection. As detailed in the journal \textit{Stroke} \cite{koge2022internal}, this system categorizes blood vessel morphologies into four empirical types: Straight, Tortuous, Coiled, and Kinked \cite{Weibel1965}.}
    \label{fig:clinical_types}
    \end{figure}

\IEEEPARstart{T}{he} morphological quantification of the extracranial internal carotid artery (ICA-C1) is clinically relevant to neurointerventional procedures, including mechanical thrombectomy for stroke treatment \cite{bouthillier1996segments,wolman2022anatomy,Baz2021,goyal2016endovascular,nogueira2018thrombectomy,pancaldi2012blood}. Traditionally, physicians assess blood vessel tortuosity through qualitative visual grading. Previous studies have classified ICA tortuosity into straight, tortuous, coiled, and kinked types, as illustrated in Fig.~\ref{fig:clinical_types} \cite{Weibel1965,koge2022internal}. Coiled and kinked blood vessels have been associated with greater difficulty in catheter navigation, lower first-pass success rates, and increased procedural risk \cite{koge2022internal,oh2022vitro}. However, visual grading depends on clinical experience and lacks clear mathematical boundaries, which limits reproducibility, automated morphological classification, and quantitative comparison.

\begin{figure}[!t]
        \centering
        \includegraphics[width=\linewidth]{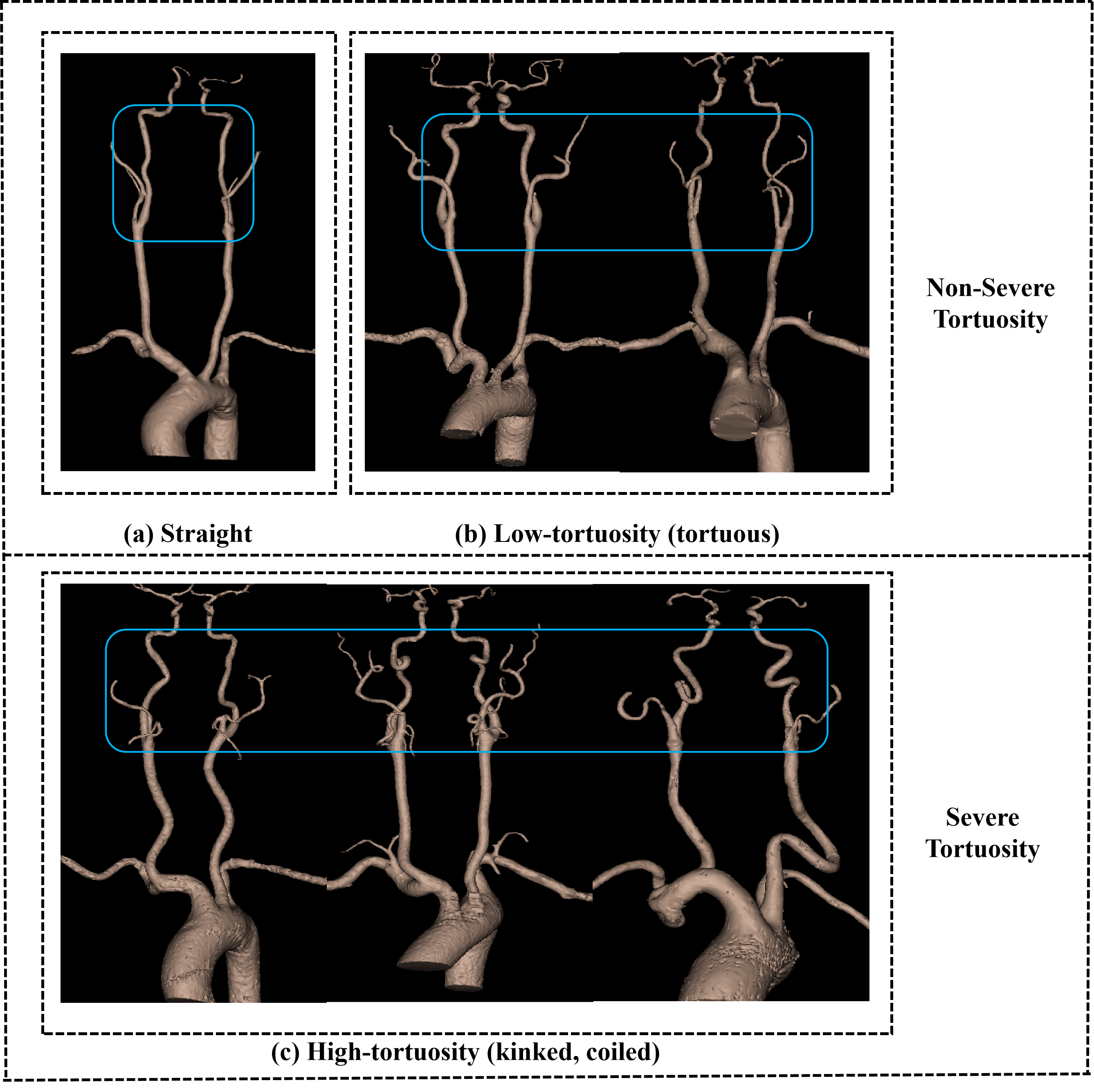} 
        \caption{The proposed geometry-driven dual-task classification framework for the ICA-C1 segment. 
        \textbf{(1) Binary Task:} The top row represents the ``Non-Severe Tortuosity'' risk category, while the bottom row represents the ``Severe Tortuosity'' risk category. 
        \textbf{(2) Ternary Task:} The ternary task uses three morphotypes: (a) Straight; (b) Low-tortuosity; and (c) High-tortuosity. 
        (3D blood vessel models are reconstructed using MIMICS software).}
        \label{fig:vessel_classification}
    \end{figure}

To operationalize this clinical consensus, we propose a geometry-driven dual-task classification framework (see Fig.~\ref{fig:vessel_classification}). Addressing the primary clinical imperative, we first introduce a binary risk screening task to rapidly differentiate severe tortuosity vessels from non-severe cases. Second, to provide a finer geometric resolution, we propose a detailed three-class morphological classification task: straight, low-tortuosity, and high-tortuosity. The binary and ternary tasks provide two related descriptions of blood vessel morphology, but their labels are analyzed separately and are not treated as an exact one-to-one mapping. Extreme localized deformations (coiled, kinked and some severe tortuosity) are consolidated into this unified ``high-tortuosity'' category. This consolidation is not only justified by the aforementioned clinical evidence, but also has computational advantages, as merging these variations reduces classification ambiguity and improves model stability. Conventionally tortuous vessels lacking extreme deformations are simply designated as ``low-tortuosity''. We emphasize that this framework serves strictly as an objective morphological stratification tool rather than a predefined linear scale of clinical severity.

Historically, the distance-based Tortuosity Index ($\mathcal{TI}$) has been a widely used macroscopic metric \cite{Bullitt2003Measuring,johnson2007robust}. However, relying on a single scalar to evaluate a complex 3D trajectory fails to capture out-of-plane twisting.\cite{Grisan2008Tortuosity,AghamohamadianSharbaf2016Tortuosity} To address this, we introduce a data-driven hybrid framework. First, a discrete geometry approach is utilized to extract 13 geometric features, bridging macroscopic integral metrics (e.g., $\mathcal{TI}$) with microscopic differential invariants directly from the centerlines \cite{an2011geometric,zhang2021application}. Second, Information Gain and Spearman correlation analyses reduce redundancy among these extracted features, yielding a selected set of 6 geometric features ($\mathcal{TI}$, $\mathcal{AC}$, $\mathcal{TC}$, $\mathcal{AC}/\mathcal{AT}$, $\mathcal{AT}$, and $\mathcal{TT}$). Curvature captures blood vessel bending, while torsion provides complementary information on non-planar twisting \cite{mardia1999estimation,blankenburg2016parameter}. Finally, the Random Forest (RF) classifier is applied to the selected geometric features to distinguish the blood vessel categories. Applying this framework, the binary classification task achieves a Macro-F1 score of 0.9206. Subsequently, for the detailed three-class morphological stratification, the RF model achieves a Macro-F1 score of 0.8626.


 The remainder of this paper is organized as follows. Section II introduces the geometric background, including the Frenet frame, curvature, torsion, and the limitations of the Tortuosity Index. Section III describes data acquisition, blood vessel reconstruction, geometric feature measurement, feature selection, machine learning models, and validation. Section IV presents the feature-selection results, statistical comparisons, classification performance, feature ablation, and the Morphological Risk Index. Section V discusses the main findings and limitations. Section VI concludes the paper and outlines future work.


\section{Background}
\label{sec:Background}

\subsection{The Frenet Frame, Curvature and Torsion}

Mathematically, a vascular centerline can be represented by a spatial curve, which is defined by a differentiable mapping $\mathbf{r}(t) = (x(t), y(t), z(t))$, where $t$ denotes a general parameter. To precisely determine and describe the geometric properties of the curve, we introduce the concept of the Frenet coordinate frame \cite{Puel1992}. The Frenet frame is a moving orthogonal system consisting of three unit basis vectors: the tangent $\mathbf{T}$, the principal normal $\mathbf{N}$, and the binormal $\mathbf{B}$, constructed directly from $\mathbf{r}(t)$, defined as:
\begin{equation}
\begin{aligned}
\mathbf{T}(t) &= \frac{\mathbf{r}'(t)}{\|\mathbf{r}'(t)\|}, \\ 
\mathbf{N}(t) &= \frac{\mathbf{T}'(t)}{\|\mathbf{T}'(t)\|}, \\ 
\mathbf{B}(t) &= \mathbf{T}(t) \times \mathbf{N}(t)
\end{aligned}
\label{eq:frenet_frame}
\end{equation}

\begin{figure}[!t]
    \centering
    \subfloat[]{\includegraphics[width=0.85\linewidth]{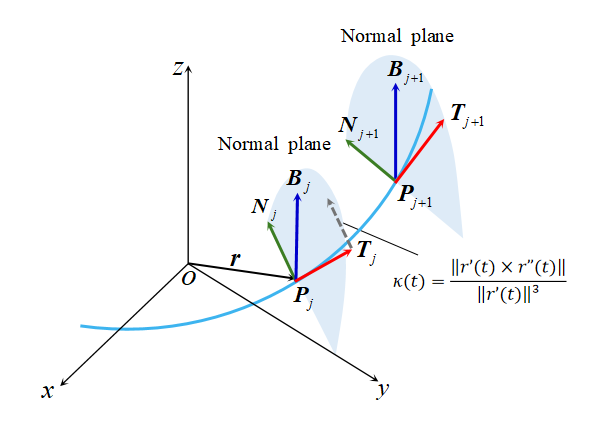}}\\[2ex]
    
    \subfloat[]{\includegraphics[width=0.85\linewidth]{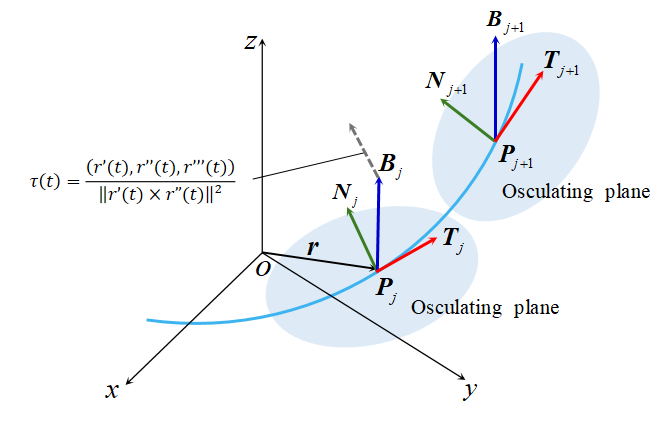}}
    
    \caption{Schematic representation of intrinsic geometric descriptors. (a) Curvature, which quantifies the local change in tangent direction within the osculating plane. (b) Torsion, which governs the out-of-plane twisting behavior, measuring the rate at which the curve deviates from its osculating plane.}
    \label{fig:geometry_schematic}
\end{figure}

Under standard smoothness conditions and where curvature is nonzero, a space curve is determined up to a rigid motion by its curvature ($\kappa$) and torsion ($\tau$). While $\kappa$ quantifies the local change in tangent direction within the osculating plane (as schematically illustrated in Fig.~\ref{fig:geometry_schematic}(a)), $\tau$ governs the non-planar twisting behavior outside the osculating plane (see Fig.~\ref{fig:geometry_schematic}(b)) \cite{zhang2021application, mardia1999estimation}. 

Parametrized by $t$ (e.g., cumulative chord length), the computational formula for 3D curvature $\kappa(t)$ is given by:
\begin{equation}
\kappa(t) = \frac{\|\mathbf{r}'(t) \times \mathbf{r}''(t)\|}{\|\mathbf{r}'(t)\|^3}
\label{eq:curvature_param}
\end{equation}

Similarly, the computational formula for the 3D torsion $\tau(t)$ is given by:
\begin{equation}
\tau(t) = \frac{(\mathbf{r}'(t),\mathbf{r}''(t),\mathbf{r}'''(t))}{\|\mathbf{r}'(t) \times \mathbf{r}''(t)\|^2}
\label{eq:torsion_param_final}
\end{equation}

\subsection{Traditional Morphological Metric: Tortuosity Index}
\label{subsec:Traditional_TI}
In previous clinical studies, researchers have conventionally used macroscopic distance metrics to preliminarily assess vascular tortuosity. The Tortuosity Index ($\mathcal{TI}$) remains a widely used measure, although recent comparative work has emphasized that different curvature- and distance-based metrics capture different aspects of vascular geometry \cite{Bullitt2003Measuring,johnson2007robust,Kashyap2022Accuracy}, defined as:
\begin{equation}
    \mathcal{TI} = \frac{\mathcal{L}}{\mathcal{D}}
    \label{eq:TI_definition}
\end{equation}

where $\mathcal{L}$ represents the total path length of the vascular centerline, and $\mathcal{D}$ denotes the straight-line Euclidean distance between its two endpoints.

While straightforward to compute, relying exclusively on $\mathcal{TI}$ introduces two critical mathematical limitations:

\subsubsection{Limitation 1: Geometric Ambiguity}
Relying on a single scalar ratio creates geometric ambiguity. $\mathcal{TI}$ inherently fails to differentiate between simple planar bending and complex spatial twisting. We prove this mathematically by constructing two analytical vascular phantoms.

\begin{itemize}[leftmargin=*, label=$\bullet$]
    \item \textbf{The Salkowski Vascular Phantom:} The spatial trajectory is defined by the following parameterized expression:
    \begin{equation}
    \resizebox{0.92\linewidth}{!}{$
    \vec{\mathbf{r}}(t)_s = \frac{n}{m} 
    \begin{pmatrix} 
        \left( \frac{(n-1)\sin((1+2n)t)}{4(1+2n)} - \frac{(1+n)\sin((1-2n)t)}{4(1-2n)} - \frac{\sin(t)}{2} \right) \hat{\mathbf{x}} \\[1.2ex]
        \left( \frac{(1-n)\cos((1+2n)t)}{4(1+2n)} + \frac{(1+n)\cos((1-2n)t)}{4(1-2n)} + \frac{\cos(t)}{2} \right) \hat{\mathbf{y}} \\[1.2ex]
        -\frac{\cos(2nt)}{4m} \hat{\mathbf{z}}
    \end{pmatrix}
    $}
    \end{equation}
    The parameters are set to $m = 1/16$ and $n = m/\sqrt{1+m^2}$, with the integration interval $t \in [0, \pi/2n]$. Based on these settings, the total length is $\mathcal{L} = 16.0000$ and the Euclidean distance is $\mathcal{D} = 8.0495$. Therefore, the Tortuosity Index is calculated as $\mathcal{TI} = \mathcal{L} / \mathcal{D} = 1.9877$.

    \item \textbf{The Helix Phantom:} The matching spatial trajectory is defined by the standard parametric equation for a cylindrical helix:
    \begin{equation}
    \vec{\mathbf{r}}(t)_h = 
    \begin{cases} 
    A\cos(t)\hat{\mathbf{x}} \\ 
    A\sin(t)\hat{\mathbf{y}} \\ 
    Bt\hat{\mathbf{z}} 
    \end{cases}
    \end{equation}
    The integration interval spans two full coils, $t \in [0, 4\pi]$, with structural parameters strictly matched as $A = 1.1003$ and $B = 0.6406$. Based on these settings, the total length is $\mathcal{L} = 16.0000$ and $\mathcal{D} = 8.0495$. Therefore, the Tortuosity Index is calculated identically as $\mathcal{TI} = \mathcal{L} / \mathcal{D} = 1.9877$. 
\end{itemize}

Both vascular phantoms yield the exact same calculation ($\mathcal{L}/\mathcal{D} = 1.9877$), yet their spatial morphologies are entirely different. This confirms that macroscopic distance metrics cannot fully capture complex structural deformations.

\subsubsection{Limitation 2: Numerical Instability}

$\mathcal{TI}$ suffers from numerical instability when evaluating extreme morphological variants. In cases where the vessel forms a tight coil or a full loop, the starting and ending points become physically proximate, causing the denominator ( $\mathcal{D}$) to approach zero. Consequently, the $\mathcal{TI}$ value artificially approaches infinity ($\mathcal{TI} \to \infty$), which may yield unstable or disproportionately large measurements \cite{brummer2020improving,johnson2007robust}.
\begin{figure}[!t]
    \centering
    \begin{subfigure}[b]{0.45\textwidth}  
        \centering
        \includegraphics[width=\linewidth]{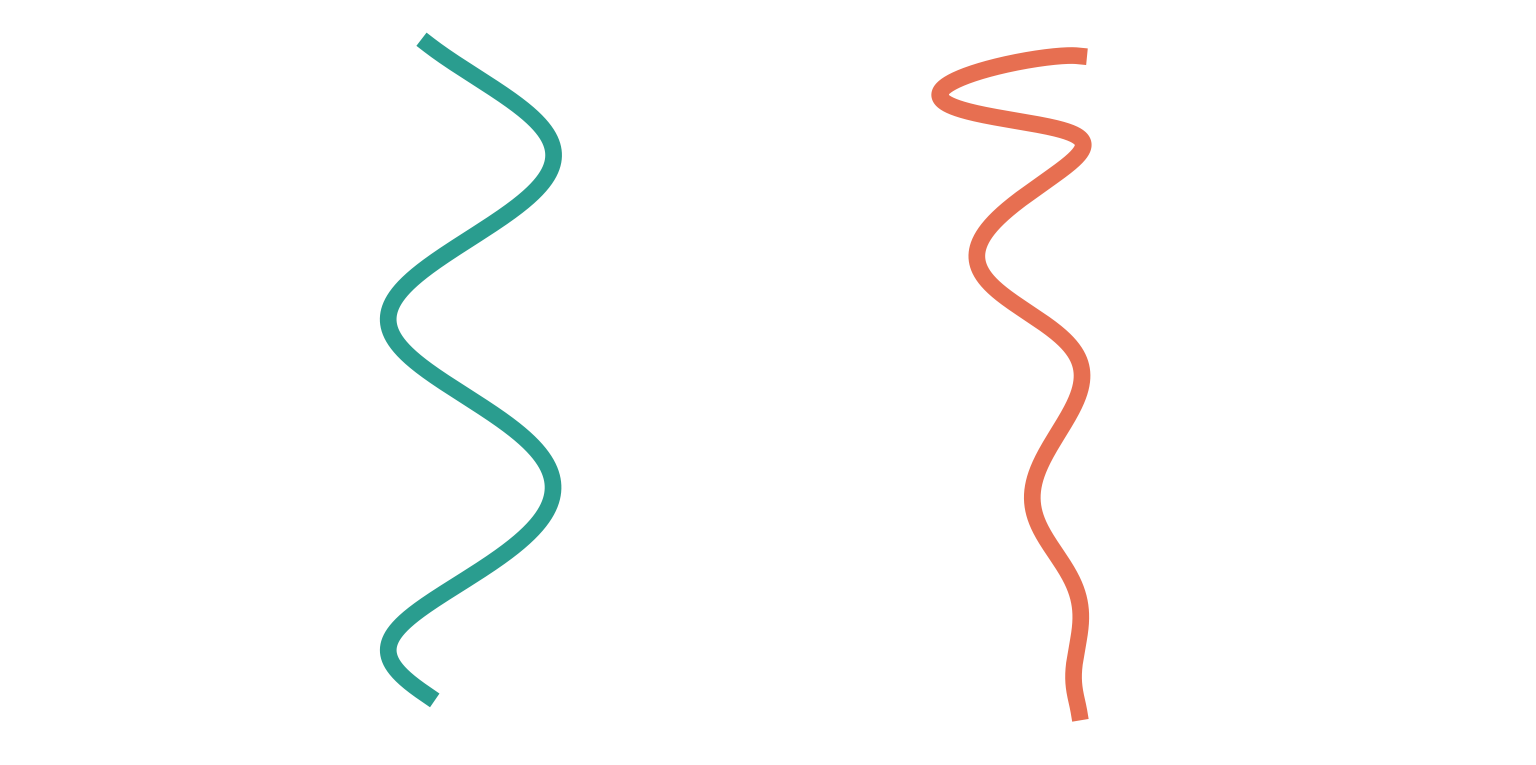} 
        \caption{Salkowski curve vs. Helix curve}
        \label{fig:ti_mathematical_curves}
    \end{subfigure}
    
    \vspace{0.3cm} 
    
    \begin{subfigure}[b]{0.45\textwidth} 
        \centering
        \includegraphics[width=\linewidth]{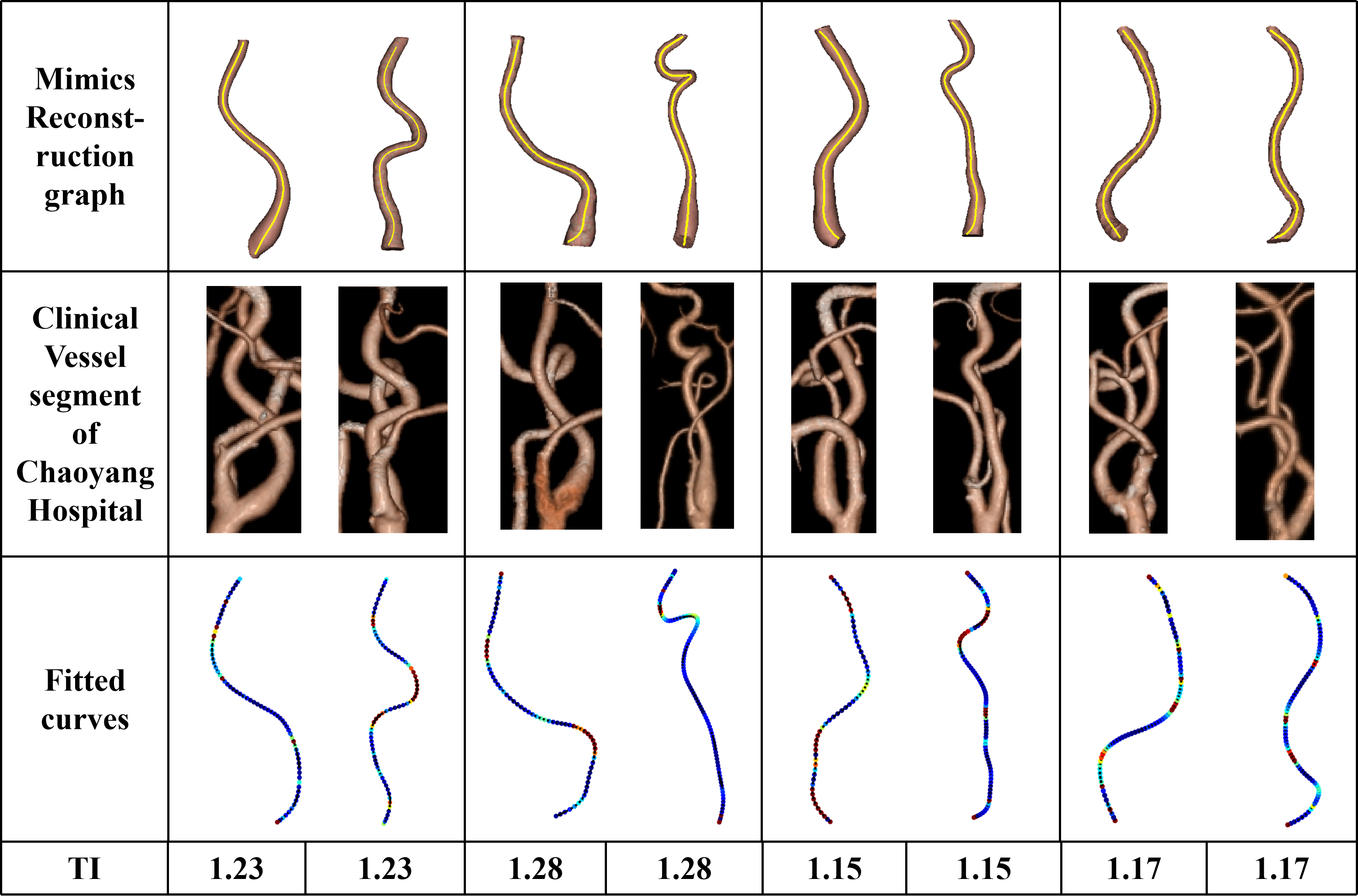} 
        \caption{Clinical ICA-C1 segments}
        \label{fig:ti_real_vessel}
    \end{subfigure}
    
    \caption{\textbf{(\subref{fig:ti_mathematical_curves})} Blood Vessel Phantoms: A Salkowski vascular phantom and a standard helix phantom yield an identical Tortuosity Index ($\mathcal{TI}$) of $1.9877$. \textbf{(\subref{fig:ti_real_vessel})} Real clinical ICA-C1 segments: Four comparative groups of patient vessels, where the two vessels within each group exhibit entirely different morphologies yet share mathematically equivalent $\mathcal{TI}$ values. Together, these visually demonstrate $\mathcal{TI}$ is insufficient to capture complex non-planar deformations.}
\label{fig:ti_ambiguity}
\end{figure}

To overcome these limitations, it is essential to introduce tortuosity metrics based on curvature and torsion. Curvature- and torsion-based features can describe non-planar deformations that are not captured by distance-based indices alone \cite{an2011geometric,zhang2021application}.

\section{Methods}
\label{sec:Methods}

\begin{figure*}[htbp]
    \centering
    \includegraphics[width=\textwidth]{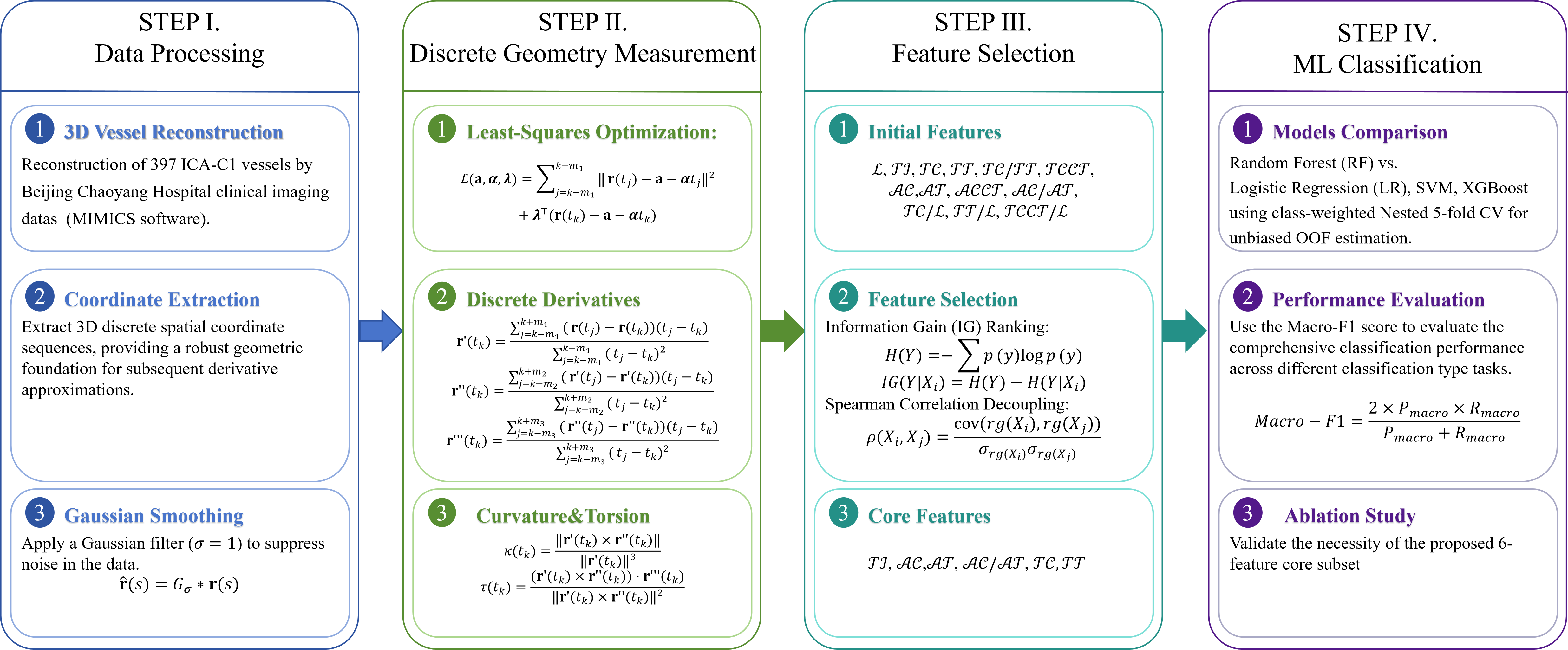}
    \caption{The overall mathematical framework and operational pipeline of the proposed method.}
    \label{fig:method_flow}
\end{figure*}

To provide a global perspective for the subsequent mathematical derivations, Fig.~\ref{fig:method_flow} outlines the overall framework. This flowchart shows the logical links among the main mathematical steps.

\subsection{Data Acquisition and Preprocessing}

\subsubsection{Clinical Data Acquisition}

The clinical data for this study were acquired from Beijing Chaoyang Hospital, Capital Medical University, between March and November 2024. With Institutional Review Board (IRB) approval, 193 subjects undergoing head and neck computed tomography angiography (CTA) were included.

\subsubsection{Vascular Reconstruction and Preprocessing}
\begin{figure*}[!t]
    \centering
    \includegraphics[width=0.9\textwidth]{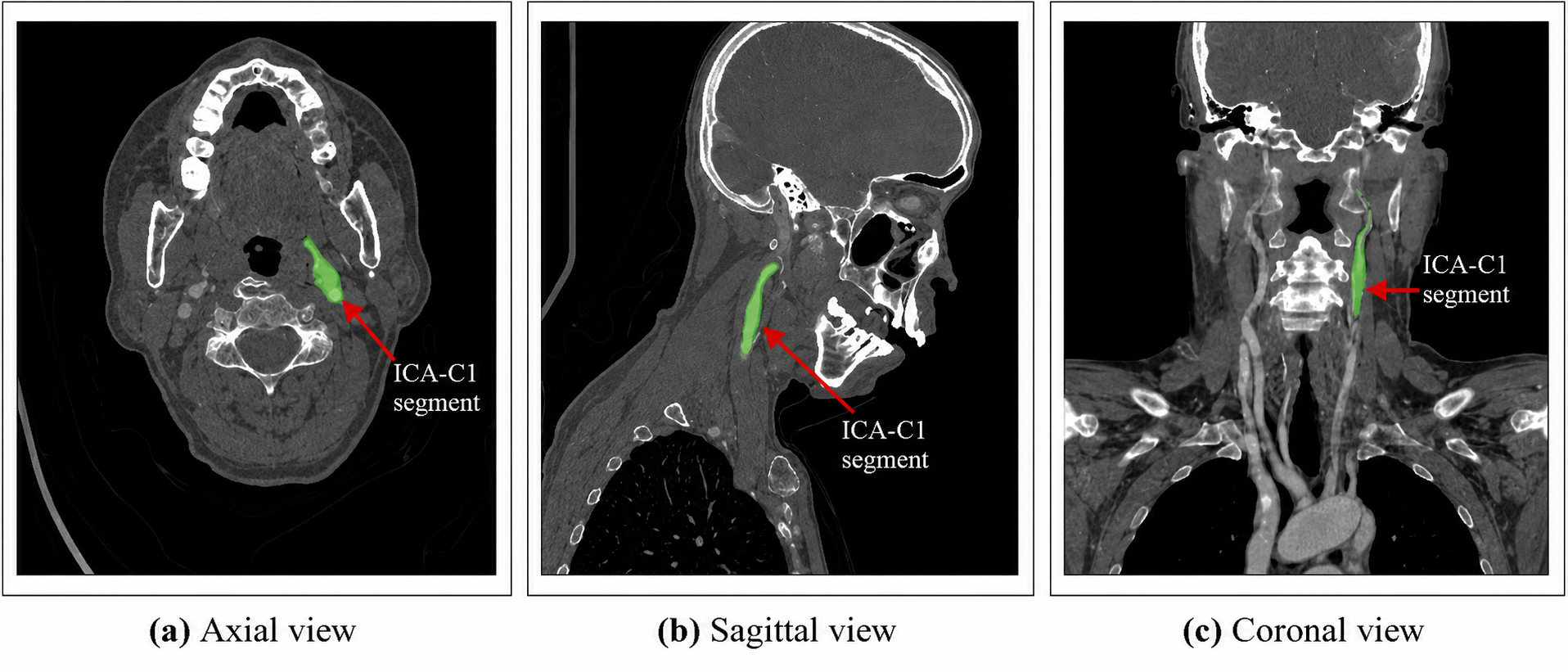} 
    \caption{Three views from Mimics medical imaging software, showing the anatomical location and morphology of the reconstructed ICA-C1 segment: (a) axial view, (b) sagittal view, and (c) coronal view. This multi-planar visualization confirms the spatial structure and precision of the vessel segment selected for morphological analysis.}
    \label{fig:mimics_three_view}
\end{figure*}

\begin{figure*}[!t] 
    \centering
    \includegraphics[width=0.95\textwidth]{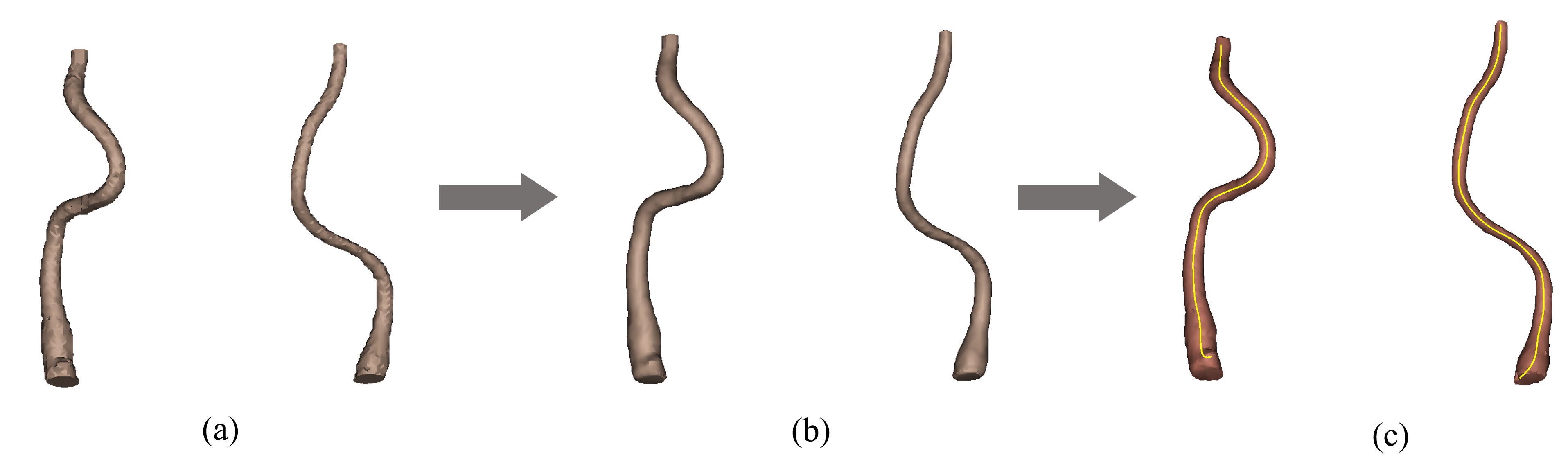} 
    \caption{Pipeline of vascular reconstruction and centerline extraction. (a) Raw vessel model with segmentation noise. (b) Smoothed vascular geometry. (c) Discrete spatial coordinates ($\mathbf{r}(t_k)$) extracted along the centerline.}
    \label{fig:mimics_pipeline}
\end{figure*}

Vascular geometric models were reconstructed via a semi-automated pipeline, following the general principles of image-based vascular modeling and centerline extraction \cite{Decroocq2023Modeling,Piccinelli2009,antiga2008image,frangi1998multiscale,aylward2002initialization,lesage2009review}. Initially, the base 3D vascular models were extracted using Mimics (Materialise, Belgium). The precise anatomical location and spatial morphology of the targeted ICA-C1 segment were confirmed through multi-planar visualizations (see Fig.~\ref{fig:mimics_three_view}). Subsequently, the raw models were smoothed in Geomagic Wrap (3D Systems, USA) and re-imported into Mimics for automated centerline extraction. Following rigorous quality control, 379 valid ICA-C1 target segments were included in the final analysis, obtaining the discrete spatial coordinate sequences $\mathbf{r}(t_k)$ with an average sampling density of approximately 0.994 points/mm (see Fig.~\ref{fig:mimics_pipeline}).

During the aforementioned centerline extraction process, the obtained discrete spatial coordinates inevitably contain noise. A standard Gaussian smoothing filter ($\sigma = 1$) is applied directly to the sequence of extracted 3D coordinates. This preprocessing reduces microscopic coordinate jitters and limits noise amplification before the extraction of geometric invariants.

\subsection{Measurement of Blood Vessel features via Discrete Geometry Method}

\subsubsection{Discrete Derivative Approximation via Lagrange Multipliers}

To calculate the curvature and torsion of the vessel centerline, we must first compute the spatial derivatives from the discrete coordinate points. Traditional finite difference methods (such as 5-point or 7-point differences) use a fixed window size. This may cause large errors and amplify noise, especially when computing the third-order derivatives. To address this problem, we adopt a discrete geometry method to estimate the derivatives \cite{an2011geometric,an2020geometric}. The main advantages of this method are its adjustable window size and noise robustness, which improve the stability of high-order derivative estimation.

\begin{figure}[!t]
    \centering
    \includegraphics[width=\linewidth]{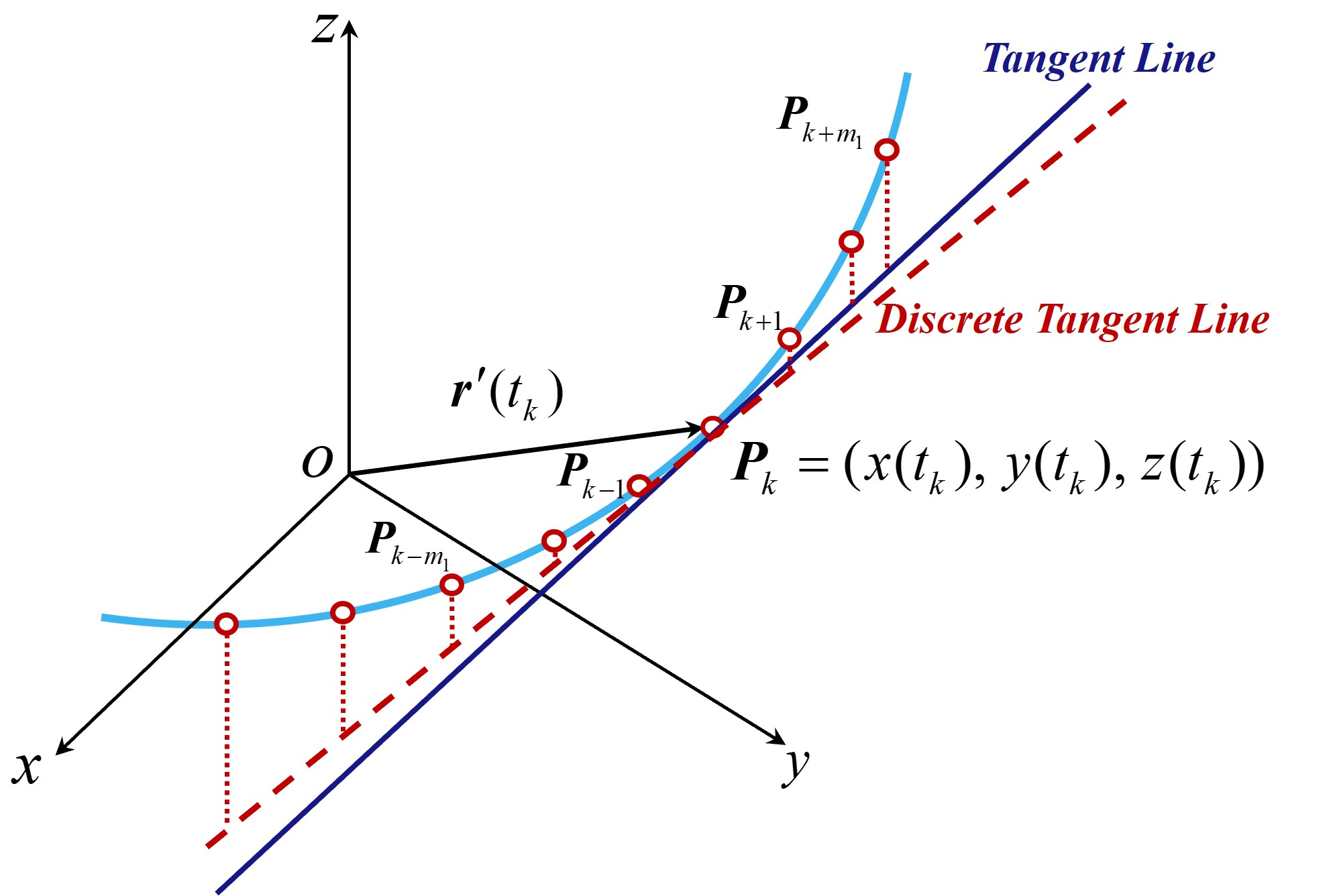}
    \caption{Sliding-window least-squares estimation of $\mathbf{r}'(t_k)$. The \textbf{Discrete Tangent Line} strictly passes through $P_k$. It is computed by minimizing the squared distances to all neighboring points within the window $V(P_k)$. The \textbf{Tangent Line} represents the exact derivative of the continuous vascular curve. As the window width shrinks to zero, the discrete tangent line converges to this true tangent line.}
    \label{fig:discrete_geometry}
\end{figure}

We denote the sliding-window radii for the first, second, and third derivatives by $\mathbf{m} = (m_1, m_2, m_3)$. In this study, $\mathbf{m}=(1,1,1)$, so $m_1=m_2=m_3=1$. The derivation below uses the first-order derivative as an example, hence the window radius $m_1$; the accompanying figure also illustrates this first-order computation (Fig.~\ref{fig:discrete_geometry}).

Unlike standard analytical curves parameterized by arc length $s$, discrete vascular centerlines extracted from medical images cannot be perfectly parameterized by $s$ without inducing approximation errors. Thus, we adopt the normalized cumulative chord length as the general parameter $t$, defined for the $i$-th point as:
\[
t_i = \frac{\sum_{j=0}^{i-1} \|\mathbf{r}_{j+1} - \mathbf{r}_j\|}{\sum_{j=0}^{N-2} \|\mathbf{r}_{j+1} - \mathbf{r}_j\|}, \qquad i=0,\ldots,N-1.
\]

Let $\mathbf{r}(t_i) \in \mathbb{R}^3$ be the position of the $i$-th centerline point. For a point $P_k$ at $t_k$, the local window of radius $m_1$ is defined as:
\begin{equation}
V(P_k) = \{\mathbf{r}(t_j) \mid k-m_1 \le j \le k+m_1\}.
\end{equation}

We construct a local linear approximation $\mathbf{r}(t) = \mathbf{a} + \boldsymbol{\alpha} t$ that passes exactly through $P_k$ and minimizes the sum of squared distances to all points within $V(P_k)$:
\begin{equation}
    \min_{\mathbf{a},\boldsymbol{\alpha}} \sum_{j=k-m_1}^{k+m_1} \|\mathbf{r}(t_j) - (\mathbf{a} + \boldsymbol{\alpha} t_j)\|^2
    \quad\mathrm{s.t.}\quad \mathbf{a} + \boldsymbol{\alpha} t_k = \mathbf{r}(t_k).
    \label{eq:optimization}
\end{equation}

In order to solve this problem, we introduce a Lagrange multiplier $\boldsymbol{\lambda}$, the Lagrangian is formulated as:
\begin{equation}
    \mathcal{L}(\mathbf{a}, \bm{\alpha}, \bm{\lambda}) =
    \sum_{j=k-m_1}^{k+m_1} \|\mathbf{r}(t_j) - \mathbf{a} - \bm{\alpha} t_j\|^2
    + \bm{\lambda}^{\top} \bigl(\mathbf{r}(t_k) - \mathbf{a} - \bm{\alpha} t_k\bigr).
    \label{eq:lagrangian}
\end{equation}

Setting the gradients with respect to $\mathbf{a}$, $\boldsymbol{\alpha}$, and $\boldsymbol{\lambda}$ to zero yields the optimal direction vector, which serves as the discrete first derivative:
\begin{equation}
\mathbf{r}'(t_k) = 
\frac{\sum_{j=k-m_1}^{k+m_1} (\mathbf{r}(t_j)-\mathbf{r}(t_k))(t_j-t_k)}
{\sum_{j=k-m_1}^{k+m_1} (t_j-t_k)^2}.
\label{eq:first_derivative}
\end{equation}

The same local constrained least-squares method is applied recursively with the corresponding window radii $m_2$ and $m_3$ to obtain the second and third derivatives:
\begin{equation}
\mathbf{r}''(t_k) = 
\frac{\sum_{j=k-m_2}^{k+m_2} (\mathbf{r}'(t_j)-\mathbf{r}'(t_k))(t_j-t_k)}
{\sum_{j=k-m_2}^{k+m_2} (t_j-t_k)^2},
\label{eq:second_derivative}
\end{equation}
\begin{equation}
\mathbf{r}'''(t_k) = 
\frac{\sum_{j=k-m_3}^{k+m_3} (\mathbf{r}''(t_j)-\mathbf{r}''(t_k))(t_j-t_k)}
{\sum_{j=k-m_3}^{k+m_3} (t_j-t_k)^2}.
\label{eq:third_derivative}
\end{equation}

Finally, the discrete curvature $\kappa(t_k)$ and torsion $\tau(t_k)$ are obtained by substituting these approximated derivatives into the fundamental differential formulas \cite{zhang2021application, brummer2020improving}:
\begin{equation}
    \kappa(t_k) = \frac{\|\mathbf{r}'(t_k) \times \mathbf{r}''(t_k)\|}
    {\|\mathbf{r}'(t_k)\|^3},
    \label{eq:curvature}
\end{equation}
\begin{equation}
    \tau(t_k) = \frac{(\mathbf{r}'(t_k) \times \mathbf{r}''(t_k)) \cdot \mathbf{r}'''(t_k)}
    {\|\mathbf{r}'(t_k) \times \mathbf{r}''(t_k)\|^2}.
    \label{eq:torsion}
\end{equation}

\subsubsection{Blood Vessel Tortuosity Feature Set}

Based on the discrete curvature and torsion estimated via the local derivative operators \cite{an2011geometric,coeurjolly2004estimating}, we establish a new set of blood vessel tortuosity features. This feature set serves as a supplement to the traditional Tortuosity Index ($\mathcal{TI}$). By capturing localized bending and twisting, these features are designed to address the limitations of global distance-based measurements \cite{Bullitt2003Measuring,Piccinelli2009,brummer2020improving,zhang2021application}. The continuous mathematical formulations for these 13 geometric features are systematically detailed in Table~\ref{tab:morphological_features}.

\begin{itemize}[leftmargin=*, label=$\bullet$]
    \item \textbf{Global Features}: Motivated by existing global vascular tortuosity measurements and recent comparisons of alternative tortuosity metrics \cite{Bullitt2003Measuring,johnson2007robust,Kashyap2022Accuracy}, we define features that summarize the macroscopic extent of bending and twisting. This category explicitly includes Total Centerline Length ($\mathcal{L}$), Tortuosity Index ($\mathcal{TI}$), Total Curvature ($\mathcal{TC}$), Total Absolute Torsion ($\mathcal{TT}$), Total Curvature-Torsion Ratio ($\mathcal{TC}/\mathcal{TT}$), and Total Combined Curvature and Torsion ($\mathcal{TCCT}$).

    \item \textbf{Averaged Features}: Motivated by curvature- and torsion-based vascular measurements \cite{brummer2020improving}, we define averaged features to characterize local geometric tendencies independently of absolute segment length. This category comprises Average Curvature ($\mathcal{AC}$), Average Absolute Torsion ($\mathcal{AT}$), Average Combined Curvature and Torsion ($\mathcal{ACCT}$), and the Average Curvature-Absolute Torsion Ratio ($\mathcal{AC}/\mathcal{AT}$).\cite{Grisan2008Tortuosity,AghamohamadianSharbaf2016Tortuosity}

    \item \textbf{Length-Normalized Features}: Following the general use of normalized vascular geometry descriptors \cite{Bogunovi2012,Lorthois2014}, we define scale-adjusted measurements, including Length-Normalized Total Curvature ($\mathcal{TC}/\mathcal{L}$), Length-Normalized Total Absolute Torsion ($\mathcal{TT}/\mathcal{L}$), and Length-Normalized TCCT ($\mathcal{TCCT}/\mathcal{L}$). 
\end{itemize}

\begin{table*}[htbp]
    \centering
    \renewcommand{\arraystretch}{2.2}
    \setlength{\tabcolsep}{8pt} 
    
    \caption{Analytical Categorization and Computational Formula of the Blood Vessel Tortuosity Features}
    \label{tab:morphological_features}
  \resizebox{\textwidth}{!}{   
    \begin{tabular}{l l c l}
    \toprule
    \textbf{Category} & \textbf{Blood Vessel Tortuosity Features} & \textbf{Symbol} & \textbf{Computational Formula} \\
    \midrule
    
    \multirow{6}{*}{\textbf{Global Metrics}} 
    & Total Centerline Length & $\mathcal{L}$ & $\int_{0}^{\mathcal{L}} ds = \int_{a}^{b} \|\mathbf{r}'(t)\| dt$ \\
    & Tortuosity Index & $\mathcal{TI}$ & $\frac{\int_{0}^{\mathcal{L}} ds}{\|\mathbf{r}(\mathcal{L}) - \mathbf{r}(0)\|} = \frac{\int_{a}^{b} \|\mathbf{r}'(t)\| dt}{\|\mathbf{r}(b) - \mathbf{r}(a)\|}$ \\
    & Total Curvature & $\mathcal{TC}$ & $\int_{0}^{\mathcal{L}} \kappa(s) ds = \int_{a}^{b} \kappa(t) \|\mathbf{r}'(t)\| dt$ \\
    & Total Absolute Torsion & $\mathcal{TT}$ & $\int_{0}^{\mathcal{L}} |\tau(s)| ds = \int_{a}^{b} |\tau(t)| \|\mathbf{r}'(t)\| dt$ \\
    & Total Curvature-Absolute Torsion Ratio & $\mathcal{TC}/\mathcal{TT}$ & $\frac{\int_{0}^{\mathcal{L}} \kappa(s) ds}{\int_{0}^{\mathcal{L}} |\tau(s)| ds} = \frac{\int_{a}^{b} \kappa(t) \|\mathbf{r}'(t)\| dt}{\int_{a}^{b} |\tau(t)| \|\mathbf{r}'(t)\| dt}$ \\
    & Total Combined Curvature \& Torsion & $\mathcal{TCCT}$ & $\int_{0}^{\mathcal{L}} \sqrt{\kappa^2(s) + \tau^2(s)} ds = \int_{a}^{b} \sqrt{\kappa^2(t) + \tau^2(t)} \|\mathbf{r}'(t)\| dt$ \\
    
    \midrule
    
    \multirow{4}{*}{\textbf{Averaged Metrics}} 
    & Average Curvature & $\mathcal{AC}$ & $\frac{1}{N} \sum_{j=1}^{N} \kappa_j$ \\
    & Average Absolute Torsion & $\mathcal{AT}$ & $\frac{1}{N} \sum_{j=1}^{N} |\tau_j|$ \\
    & Average Combined Curvature \& Torsion & $\mathcal{ACCT}$ & $\frac{1}{N} \sum_{j=1}^{N} \sqrt{\kappa_j^2 + \tau_j^2}$ \\
    & Average Curvature-Absolute Torsion Ratio & $\mathcal{AC}/\mathcal{AT}$ & $\frac{\sum_{j=1}^{N} \kappa_j}{\sum_{j=1}^{N} |\tau_j|}$ \\
    
    \midrule
    
    \multirow{3}{*}{\textbf{Length-Normalized Metrics}} 
    & Length-Normalized Total Curvature & $\mathcal{TC}/\mathcal{L}$ & $\frac{1}{\mathcal{L}} \int_{0}^{\mathcal{L}} \kappa(s) ds = \frac{\int_{a}^{b} \kappa(t) \|\mathbf{r}'(t)\| dt}{\int_{a}^{b} \|\mathbf{r}'(t)\| dt}$ \\
    & Length-Normalized Total Absolute Torsion & $\mathcal{TT}/\mathcal{L}$ & $\frac{1}{\mathcal{L}} \int_{0}^{\mathcal{L}} |\tau(s)| ds = \frac{\int_{a}^{b} |\tau(t)| \|\mathbf{r}'(t)\| dt}{\int_{a}^{b} \|\mathbf{r}'(t)\| dt}$ \\
    & Length-Normalized Total Combined Curvature \& Torsion & $\mathcal{TCCT}/\mathcal{L}$ & $\frac{1}{\mathcal{L}} \int_{0}^{\mathcal{L}} \sqrt{\kappa^2(s) + \tau^2(s)} ds = \frac{\int_{a}^{b} \sqrt{\kappa^2(t) + \tau^2(t)} \|\mathbf{r}'(t)\| dt}{\int_{a}^{b} \|\mathbf{r}'(t)\| dt}$ \\
    \bottomrule
    \end{tabular}
    }
    \vspace{4pt}
    \flushleft{\footnotesize \textit{Note:} $s \in [0, \mathcal{L}]$ and $t \in [a, b]$ denote the intrinsic arc-length and generalized parameterizations, respectively. In our numerical implementation, \textbf{Global} and \textbf{Length-Normalized Metrics} are robustly computed using the trapezoidal rule for integration, while \textbf{Averaged Metrics} are calculated as the arithmetic mean over $N$ valid discrete sampling points.}
\end{table*}

These extracted features form the initial dataset, which is subsequently screened by Information Gain (IG) and correlation mechanisms to reduce mathematical redundancy, yielding a selected geometric feature set for the machine learning classifiers \cite{Li2017Feature,Vergara2014Review}.

\subsection{Feature Selection via Information Gain and Correlation Analysis Methods}
The initial 13 blood vessel tortuosity features, derived primarily from curvature and torsion, inherently contain significant mathematical redundancy. To systematically decouple these variables, we implemented a cascade feature selection method. Specifically, an Information Gain (IG) criterion was first utilized to evaluate their discriminative power, followed by a Spearman correlation analysis to eliminate redundant information \cite{Li2017Feature}.

\subsubsection{Information Gain Ranking}

To capture the complex nonlinear mappings between vascular spatial distortions and morphological classifications, we employed Information Gain (IG) based on Shannon Information Theory \cite{Vergara2014Review, cover1991elements}. 

Let $Y$ represent the discrete morphological label space with $c$ categories. Its inherent uncertainty is characterized by the Shannon Entropy $H(Y)$:
\begin{equation}
    H(Y) = -\sum_{i=1}^{c} P(y_i) \log_2 P(y_i)
    \label{eq:shannon_entropy}
\end{equation}
where $P(y_i)$ is the marginal probability of category $y_i$.

For any continuous tortuosity feature $X$, empirically discretized into $K$ intervals via quantile-based binning, the Conditional Entropy $H(Y|X)$ is defined as:
\begin{equation}
    H(Y|X) = -\sum_{j=1}^{K} P(x_j) \sum_{i=1}^{c} P(y_i|x_j) \log_2 P(y_i|x_j)
    \label{eq:conditional_entropy}
\end{equation}
where $P(x_j)$ is the probability of $X$ falling into the $j$-th interval. 

The Information Gain $IG(Y,X)$ quantifies the reduction in uncertainty after observing $X$:
\begin{equation}
    IG(Y,X) = H(Y) - H(Y|X)
    \label{eq:information_gain}
\end{equation}

By calculating and sorting the IG values for all 13 tortuosity features, we established a preliminary importance ranking. This ranking directly serves as the foundational criterion to eliminate lower-ranked redundant features during the subsequent correlation analysis.

\subsubsection{Spearman's Correlation Analysis}

To mitigate multicollinearity and feature redundancy, we performed a non-parametric Spearman's rank correlation analysis following the IG screening. This method assesses monotonic relationships and is well-suited for non-normally distributed continuous data and data with relevant outliers \cite{Schober2018Correlation}.

For any two geometric variables $X_i$ and $X_j$, the Spearman's correlation coefficient $\rho$ is calculated using the covariance of their respective rank variables, $rg(X_i)$ and $rg(X_j)$:
\begin{equation}
    \rho(X_i, X_j) = \frac{\mathrm{cov}(rg(X_i), rg(X_j))}{\sigma_{rg(X_i)} \sigma_{rg(X_j)}}
    \label{eq:spearman}
\end{equation}

When highly correlated feature pairs were identified, the feature with the lower IG ranking was systematically eliminated. This sequential screening process refined the initial 13 parameters into a decoupled core set of 6 features: $\boldsymbol{\mathcal{TI}}$, $\boldsymbol{\mathcal{AC}}$, $\boldsymbol{\mathcal{TC}}$, $\boldsymbol{\mathcal{AC}}/\boldsymbol{\mathcal{AT}}$, $\boldsymbol{\mathcal{AT}}$, and $\boldsymbol{\mathcal{TT}}$. Because this subset mathematically captures both the in-plane bending and out-of-plane twisting of the vessels without redundancy, it serves as the unified input space for both the binary risk screening and ternary morphological grading tasks.

\subsection{Machine Learning-Based Vascular Tortuosity Classification}

\subsubsection{Classification Method Comparison}

Our diagnostic goal is to establish a robust mapping from six extracted core vascular tortuousness features to morphological risk categories. Before modeling, we rigorously apply standard zero-mean and unit-variance normalization to all features to eliminate mathematical bias caused by different numerical scales.

To address the imbalance in the number of morphological types in clinical data, we avoid using synthetic data augmentation (e.g., SMOTE) to preserve the true geometric distribution. Instead, we employ a cost-sensitive empirical risk minimization strategy. During training, class weighting is used for RF, LR, and SVM so that smaller classes receive more weight. The XGBoost model is trained without an additional class-weight term.

With the dataset standardized and the empirical distribution preserved, we evaluate four different classification algorithms:
\begin{itemize}
    \item \textbf{Logistic Regression (LR):} Modeled as the linear baseline. To evaluate feature synergy and constrain structural risk in the binary screening task, the objective function was optimized with an $L_1$ (Lasso) penalty to induce sparsity:
    \begin{equation}
        -\frac{1}{N} \sum_{i=1}^{N} \left[ y_i \log \hat{y}_i + (1 - y_i) \log (1 - \hat{y}_i) \right] + \frac{1}{C} ||\mathbf{w}||_1
        \label{eq:lr_cost}
    \end{equation}
    For the ternary grading task, this was extended via a multinomial formulation utilizing a standard $L_2$ penalty to map continuous geometric transitions.

    \item \textbf{Support Vector Machine (SVM):} Employed to handle nonlinear separability by projecting the features via the Radial Basis Function (RBF) kernel, optimizing the soft-margin boundaries:
    \begin{equation}
        K(\mathbf{x}_i, \mathbf{x}_j) = \exp(-\gamma ||\mathbf{x}_i - \mathbf{x}_j||^2)
        \label{eq:svm_rbf}
    \end{equation}

    \item \textbf{eXtreme Gradient Boosting (XGBoost):} Constructed an additive tree ensemble by sequentially minimizing the residuals of prior predictions. The optimization leverages a second-order Taylor expansion of the loss function and incorporates an explicit penalty $\Omega(f_t)$ to restrict tree complexity:
    \begin{equation}
        \mathcal{L}^{(t)} \approx \sum_{i=1}^{N} \left[ g_i f_t(\mathbf{x}_i) + \frac{1}{2} h_i f_t^2(\mathbf{x}_i) \right] + \Omega(f_t)
        \label{eq:xgb_obj}
    \end{equation}

    \item \textbf{Random Forest (RF):} Operated on the bagging paradigm to mitigate high variance. It constructs $K$ decorrelated decision trees $\{h_k(\mathbf{x})\}_{k=1}^{K}$ using stochastic feature subsets, establishing the final classification via majority voting \cite{breiman2001random}:
    \begin{equation}
        \hat{y} = \text{mode} \{ h_k(\mathbf{x}) \}_{k=1}^{K}
        \label{eq:rf_vote}
    \end{equation}
\end{itemize}

\subsubsection{Model Validation and Performance}

Given the finite sample size typical of medical cohorts, we implemented a nested 5-fold cross-validation architecture to prevent hyperparameter selection bias. The inner 3-fold loop exhaustively optimized the hyperparameters via grid search (e.g., ensemble size, learning rate, and regularization strength). The optimal models were subsequently evaluated solely on the mutually exclusive outer test folds. The concatenation of these iterations formed a complete out-of-fold (OOF) prediction set, ensuring an unbiased mathematical estimation of the model's empirical risk on unseen geometric data.

The classification outcomes are fundamentally categorized into four outcomes: True Positives (TP), False Positives (FP), True Negatives (TN), and False Negatives (FN), as illustrated in the standard confusion matrix (Table~\ref{tab:confusion_matrix}).

\begin{table}[htbp]
    \centering
    \caption{Standard Confusion Matrix for Classification Analysis}
    \label{tab:confusion_matrix}
    \begin{tabular}{l cc}
    \toprule
    & \textbf{Predicted Positive} & \textbf{Predicted Negative} \\
    \midrule
    \textbf{Actual Positive} & True Positive (TP) & False Negative (FN) \\
    \textbf{Actual Negative} & False Positive (FP) & True Negative (TN) \\
    \bottomrule
    \end{tabular}
\end{table}

To ensure equitable evaluation across the highly imbalanced clinical morphotypes, model efficacy was quantified using macro-averaged metrics. Based on the fundamental outcomes defined in Table~\ref{tab:confusion_matrix}, for any given morphological class $c$ (evaluated in a One-vs-Rest manner), the Precision ($P_c$) and Recall ($R_c$) are defined as:
\begin{equation}
    P_c = \frac{TP_c}{TP_c + FP_c}, \quad R_c = \frac{TP_c}{TP_c + FN_c}
    \label{eq:class_pr}
\end{equation}

For each class $c$, the F1 score is the harmonic mean of class-specific precision and recall. The Macro-F1 score is then obtained by averaging the class-specific F1 scores across all $C$ classes \cite{sokolova2009systematic}:
\begin{equation}
    \text{Macro-F1} = \frac{1}{C}\sum_{c=1}^{C}\frac{2P_cR_c}{P_c+R_c}
    \label{eq:macro_f1}
\end{equation}

\subsubsection{Backward Ablation Study}

To assess feature redundancy and isolate the optimal diagnostic subspace, a backward feature ablation study was conducted within the nested cross-validation framework. Features were systematically eliminated, and their impact on the OOF Macro-F1 score was recorded. Acknowledging the intrinsic measurement uncertainty in physiological imaging, minor algorithmic fluctuations (e.g., at the $10^{-3}$ scale) were deemed statistically equivalent under a predefined tolerance interval. A geometric feature was eliminated only if its removal did not disrupt the nonlinear compensatory mechanisms vital for accurate morphotype classification. The removal order follows the ascending ranking of Information Gain (i.e., the feature with the lowest IG is removed first), ensuring a consistent criterion for evaluating the marginal contribution of each feature group.

\section{RESULTS}

\begin{figure*}[htbp]
    \centering
    \includegraphics[width=\textwidth]{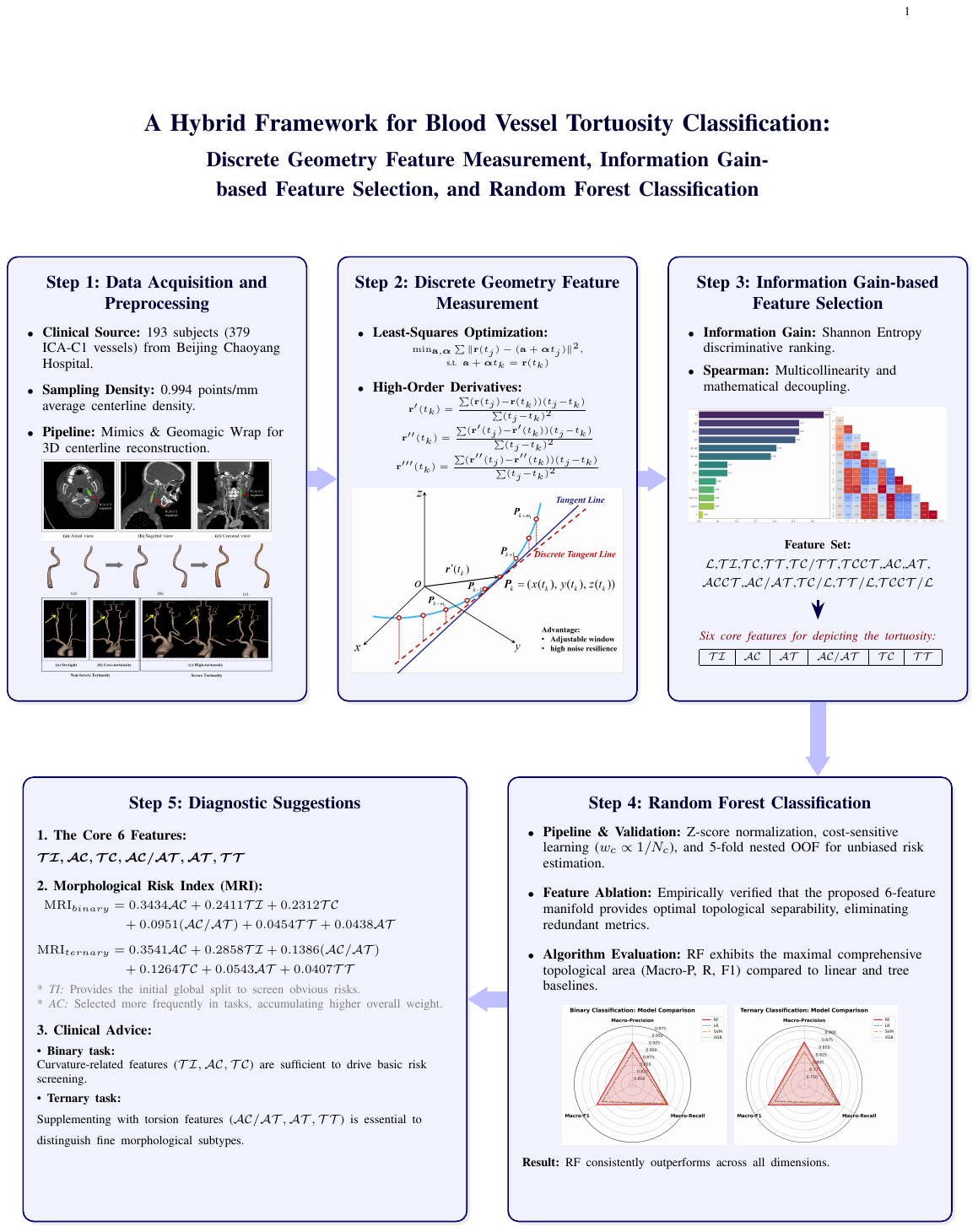}
    \caption{The overall flowchart of the proposed data-driven hybrid framework. The upper row illustrates the morphological data processing pipeline from left to right. The lower row details the S-shaped workflow transition, including the multi-model Machine Learning Classification and the final dual-task Clinical Outputs driven exclusively by the 6 optimized core geometric features.}
    \label{fig:all_result_flow}
\end{figure*}

This section systematically presents and discusses the numerical experiments and application results of the proposed mathematical model. To provide a clear overview of the global logic, Figure \ref{fig:all_result_flow} schematizes the entire derivation chain from input data to final outputs, illustrating how individual experiments validate core mathematical properties and consequently substantiate the clinical conclusions.

\subsection{Identification of Core Geometric Features}

Following the feature selection method outlined in our methodology, we sequentially applied Information Gain (IG) ranking and Spearman correlation analysis to refine the 13 initial indices into an optimal, non-redundant feature subset.

\begin{figure}[htbp]
    \centering
    \includegraphics[width=0.95\linewidth]{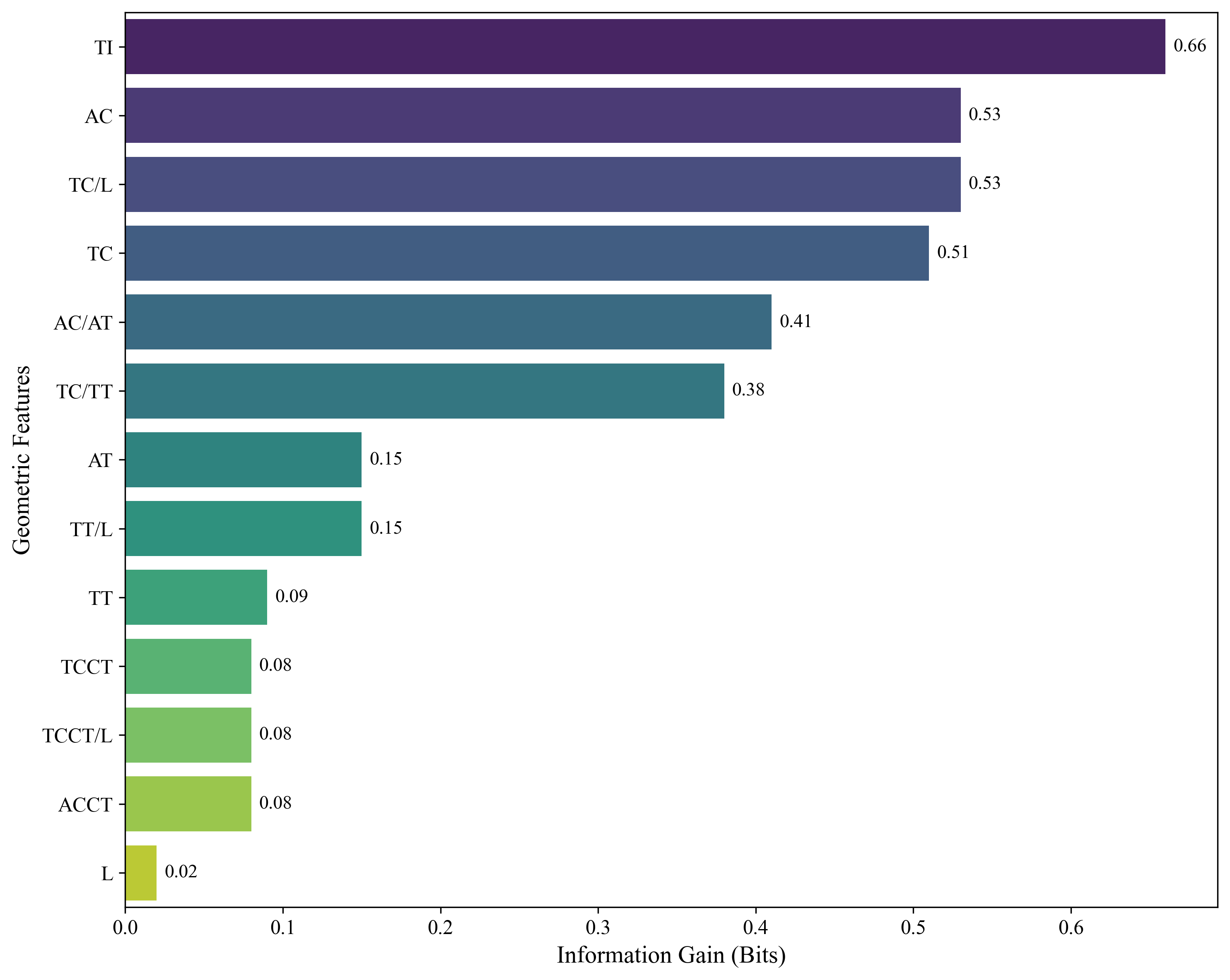} 
    \caption{Information Gain ranking of the extracted blood vessel tortuosity features. The bar chart ranks the importance of the features from the highest IG values to the lowest. The results indicate that the Tortuosity Index ($\mathcal{TI}$) has the highest IG value, followed by Average Curvature ($\mathcal{AC}$) and Length-Normalized Total Curvature ($\mathcal{TC}/\mathcal{L}$).}
    \label{fig:information_gain}
\end{figure}

First, we calculated IG values based on the physician's previous ternary morphological classification to quantify the discriminative power of each feature for classification \cite{Vergara2014Review,Shannon1948Mathematical}. As shown in Fig.~\ref{fig:information_gain}, the macroscopic tortuosity index ${\mathcal{TI}}$ obtained the highest IG value of $0.66$, while the length feature ${\mathcal{L}}$ was excluded due to its lowest ranking. However, due to its inherent geometric ambiguity (Section~\ref{subsec:Traditional_TI}), $\mathcal{TI}$ cannot independently distinguish complex morphological types. The microscopic curvature descriptors provide complementary information, achieving an IG value of 0.53 for both the average curvature ($\boldsymbol{\mathcal{AC}}$) and the length-normalized total curvature ($\boldsymbol{\mathcal{TC}}/\boldsymbol{\mathcal{L}}$). Size-related features, such as total length ($\boldsymbol{\mathcal{L}}$), rank lower due to their sensitivity to clipping boundaries.

\begin{figure}[htbp]
    \centering
    \includegraphics[width=\columnwidth]{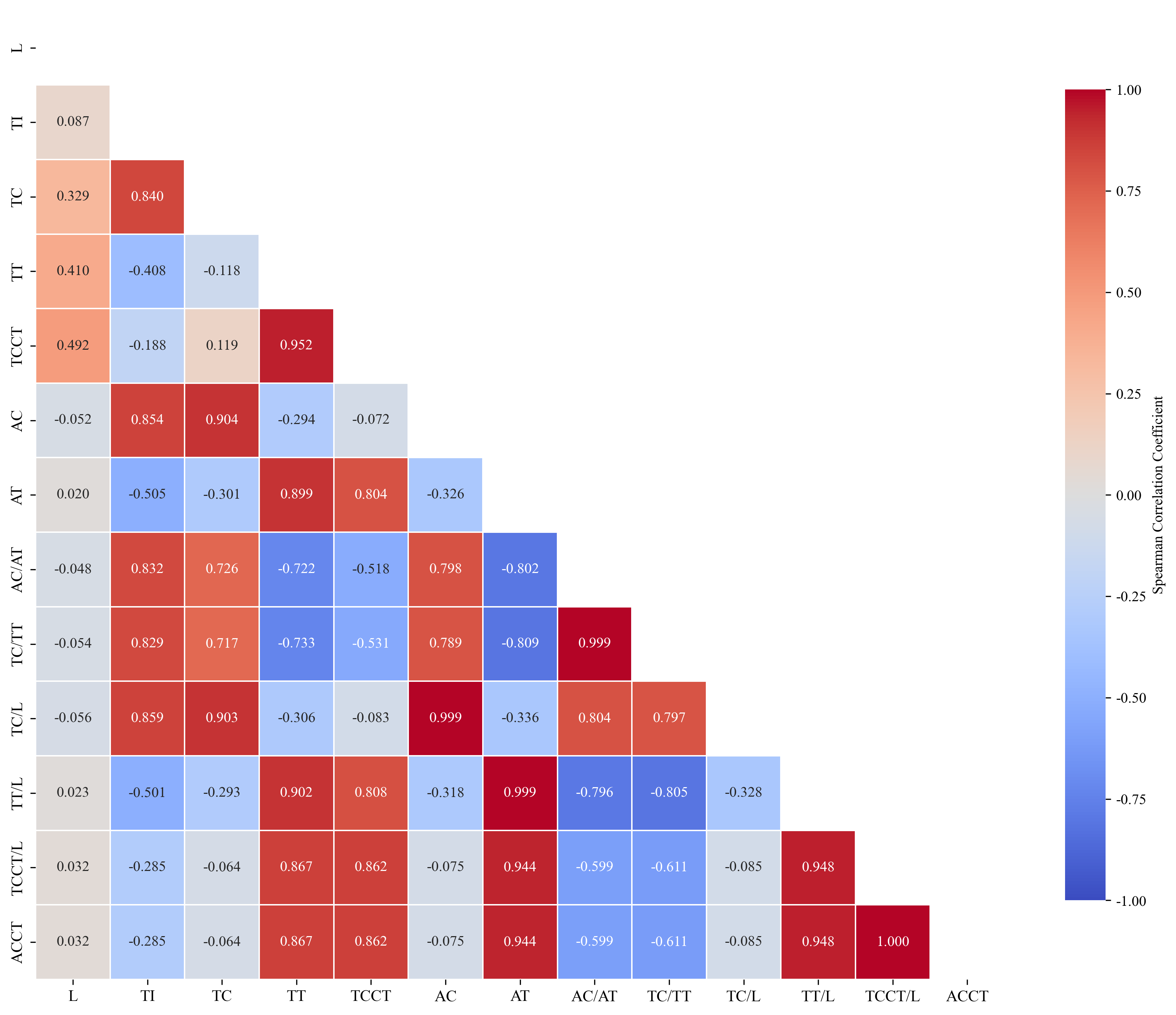} 
\caption{Spearman rank correlation matrix of the extracted morphological features. To mitigate mathematical redundancy and multicollinearity, a prespecified pruning rule was applied: among highly coupled variable pairs ($|\rho| > 0.95$), the feature with the lower Information Gain was eliminated. Based on this criterion, $\mathcal{TC}/\mathcal{L}$, $\mathcal{TC}/\mathcal{TT}$, $\mathcal{TT}/\mathcal{L}$, $\mathcal{TCCT}$, and $\mathcal{ACCT}$ were removed. The additional exclusion of $\mathcal{L}$ and $\mathcal{TCCT}/\mathcal{L}$ is explained in the text. The selection process retains 6 core geometric features with reduced redundancy ($\boldsymbol{\mathcal{TI}}$, $\boldsymbol{\mathcal{AC}}$, $\boldsymbol{\mathcal{TC}}$, $\boldsymbol{\mathcal{AC}}/\boldsymbol{\mathcal{AT}}$, $\boldsymbol{\mathcal{AT}}$, and $\boldsymbol{\mathcal{TT}}$) for subsequent model training.}
    \label{fig:spearman_matrix}
\end{figure}

Subsequently, to reduce multicollinearity among high-performance features, we evaluated their Spearman rank correlation coefficients \cite{Schober2018Correlation}. As shown in the figure (Fig.~\ref{fig:spearman_matrix}), several derived metrics exhibit near-perfect positive correlations. By applying the correlation rule ($|\rho| > 0.95$), we first removed the lower-IG feature from each highly correlated pair. This step removed $\mathcal{TC}/\mathcal{L}$, $\mathcal{TC}/\mathcal{TT}$, $\mathcal{TT}/\mathcal{L}$, $\mathcal{TCCT}$, and $\mathcal{ACCT}$. We then excluded $\mathcal{L}$ because it had the lowest IG value (0.02) and depends on the selected segment endpoints. We also excluded $\mathcal{TCCT}/\mathcal{L}$ because its IG value was low (0.08), it was perfectly correlated with $\mathcal{ACCT}$ in this dataset ($\rho=1.00$), and separate curvature and torsion features were already retained.

Ultimately, this process reduced the initial parameter set to six core features:

\begin{itemize}

\item Macroscopic feature ($\boldsymbol{\mathcal{TI}}$): Captures the overall spatial deformation and total elongation of the vessel.

\item Local bending features ($\boldsymbol{\mathcal{AC}}$ and $\boldsymbol{\mathcal{TC}}$): Quantify the local bending intensity ($\boldsymbol{\mathcal{AC}}$) and the total cumulative bending ($\boldsymbol{\mathcal{TC}}$) along the spatial curve, respectively.

\item Spatial torsion features ($\boldsymbol{\mathcal{AT}}$ and $\boldsymbol{\mathcal{TT}}$): Measure the local rate ($\boldsymbol{\mathcal{AT}}$) and its cumulative magnitude ($\boldsymbol{\mathcal{TT}}$) of out-of-plane torsion.

\item Bending-torsion coupling feature ($\boldsymbol{\mathcal{AC}}/\boldsymbol{\mathcal{AT}}$): An important ratio (IG = $0.41$) that reflects the inherent geometric interaction between curvature and torsion, serving as a key biomarker for morphological differentiation.
\end{itemize}

These six retained variables form a feature set with reduced redundancy for the subsequent machine learning classifiers.


\subsection{Statistical Analysis of Vascular Morphology}
\begin{figure*}[!p]
    \centering
    
    \captionsetup{type=table}
    \caption{Statistical Comparison of Features Across the Three Morphological Classes}
    \label{tab:statistical_comparison}
    \small  
    \resizebox{0.95\textwidth}{!}{%
    \begin{tabular}{c c c c c c c} 
    \toprule
    \textbf{Feature} & \textbf{Straight} & \textbf{Low Tortuosity} & \textbf{High Tortuosity} & \textbf{Kruskal--Wallis H} & \textbf{$p$-value} & \textbf{Effect Size ($\varepsilon^2$)} \\
    \midrule
    $\mathcal{TI}$ & $1.0561\,(1.0390, 1.0762)$ & $1.1521\,(1.1071, 1.1856)$ & $1.3137\,(1.2353, 1.4480)$ & 270.6358 & $< 0.001$ & 0.7160 \\
    $\mathcal{AC}$ & $0.0412\,(0.0360, 0.0468)$ & $0.0552\,(0.0470, 0.0598)$ & $0.0763\,(0.0678, 0.0885)$ & 245.6507 & $< 0.001$ & 0.6499 \\
    $\mathcal{TC}$ & $2.8145\,(2.4794, 3.3834)$ & $3.8490\,(3.3984, 4.2469)$ & $5.3698\,(4.8917, 6.4028)$ & 232.7869 & $< 0.001$ & 0.6158 \\
    $\mathcal{AC}/\mathcal{AT}$ & $0.1896\,(0.1409, 0.2357)$ & $0.3445\,(0.2771, 0.4336)$ & $0.5062\,(0.3934, 0.6408)$ & 193.5607 & $< 0.001$ & 0.5121 \\
    $\mathcal{AT}$ & $0.2157\,(0.1773, 0.2649)$ & $0.1551\,(0.1310, 0.1847)$ & $0.1544\,(0.1284, 0.1817)$ & 78.9389 & $< 0.001$ & 0.2088 \\
    $\mathcal{TT}$ & $15.4407\,(12.3997, 20.0973)$ & $11.2023\,(9.1529, 13.2979)$ & $11.5198\,(9.8222, 13.7029)$ & 64.0805 & $< 0.001$ & 0.1695 \\
    \bottomrule
    \end{tabular}%
    }
    \begin{flushleft}
    \footnotesize
    \textit{Note:} Values are presented as median (Q1, Q3). Sample sizes for the Straight, Low Tortuosity, and High Tortuosity groups are $N=93$, $174$, and $112$, respectively. $P$-values are derived from the Kruskal--Wallis H test, and $\varepsilon^2$ denotes the reported effect size.
    \end{flushleft}
    
    \vspace{0.5em}
    
    \captionsetup{type=figure}
    
    \begin{subfigure}{0.26\textwidth}
        \centering
        \includegraphics[width=\linewidth]{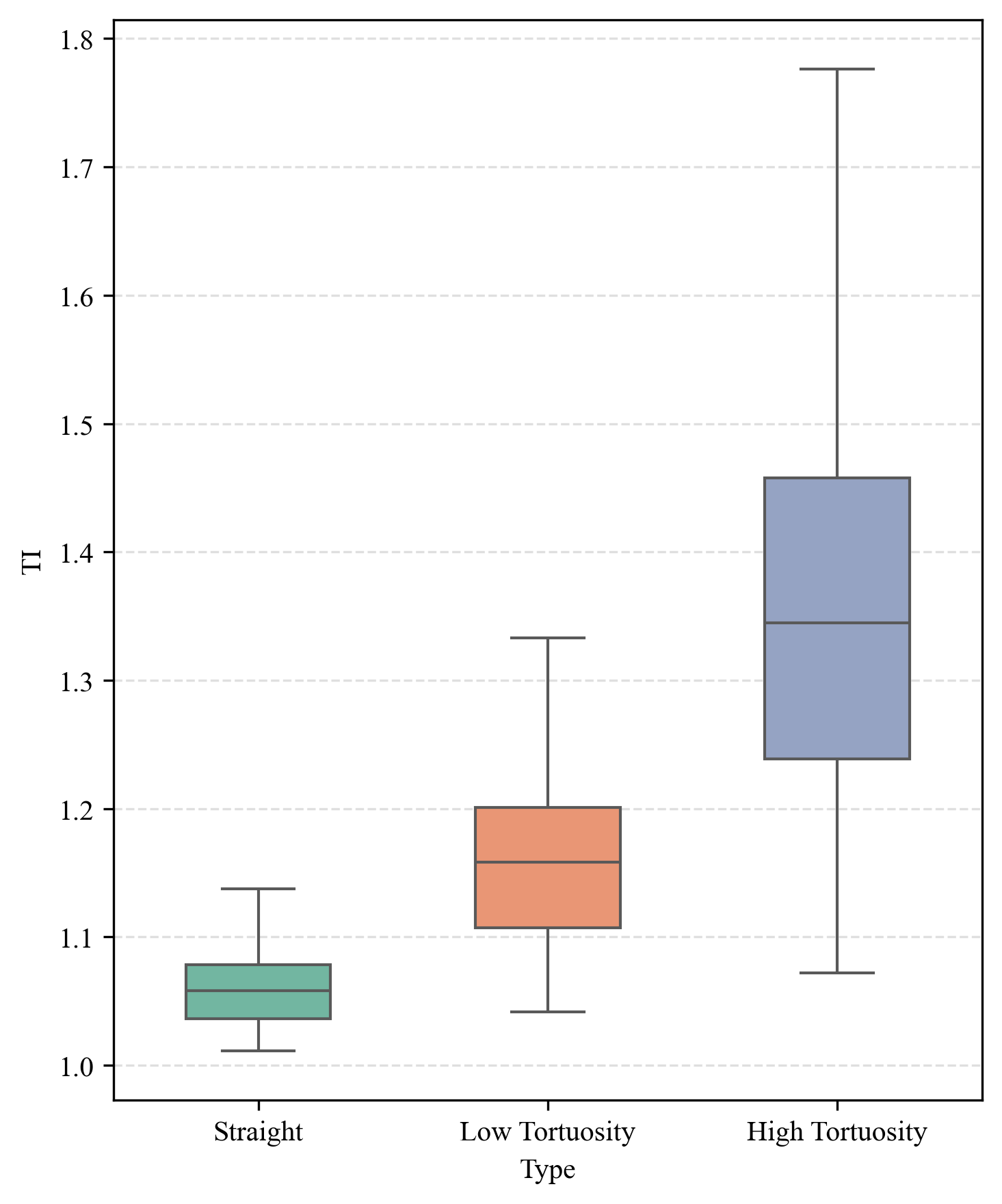} 
        \caption{$\mathcal{TI}$}
        \label{fig:ti_box}
    \end{subfigure}
    \hfill
    \begin{subfigure}{0.26\textwidth}
        \centering
        \includegraphics[width=\linewidth]{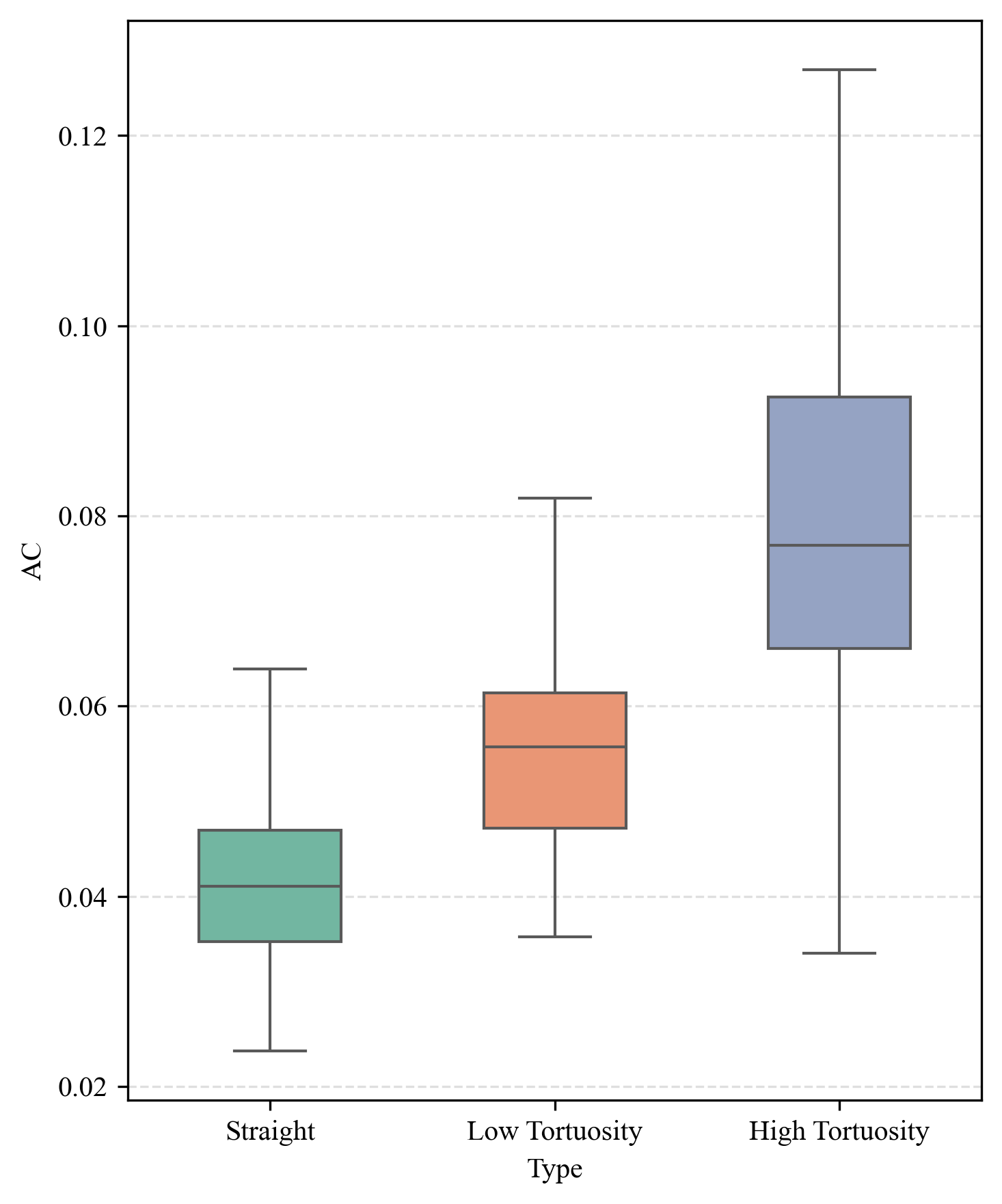} 
        \caption{$\mathcal{AC}$}
        \label{fig:ac_box}
    \end{subfigure}
    \hfill
    \begin{subfigure}{0.26\textwidth}
        \centering
        \includegraphics[width=\linewidth]{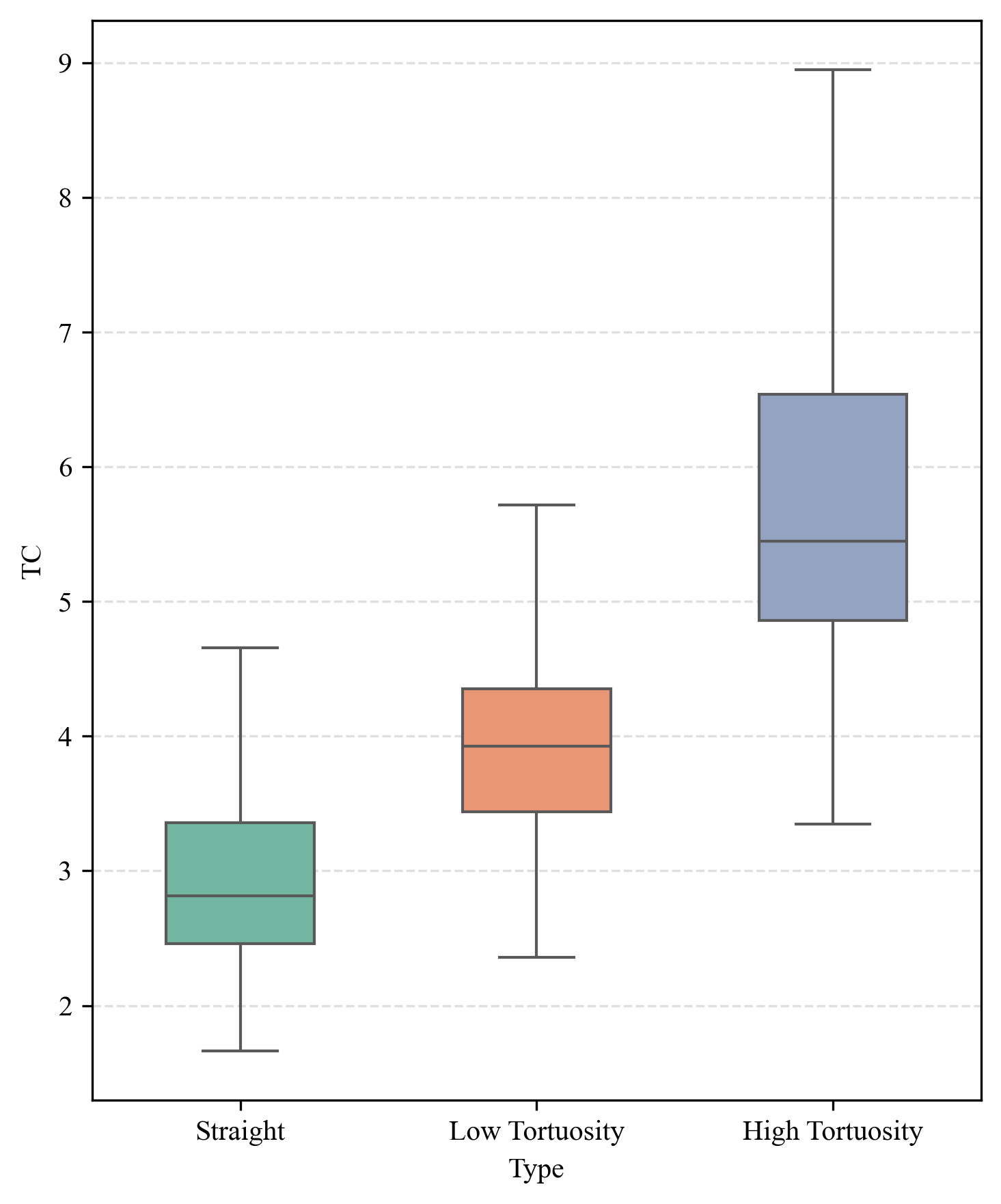} 
        \caption{$\mathcal{TC}$}
        \label{fig:tc_box}
    \end{subfigure}
    
    \vspace{0.5em}  
    
    \begin{subfigure}{0.26\textwidth}
        \centering
        \includegraphics[width=\linewidth]{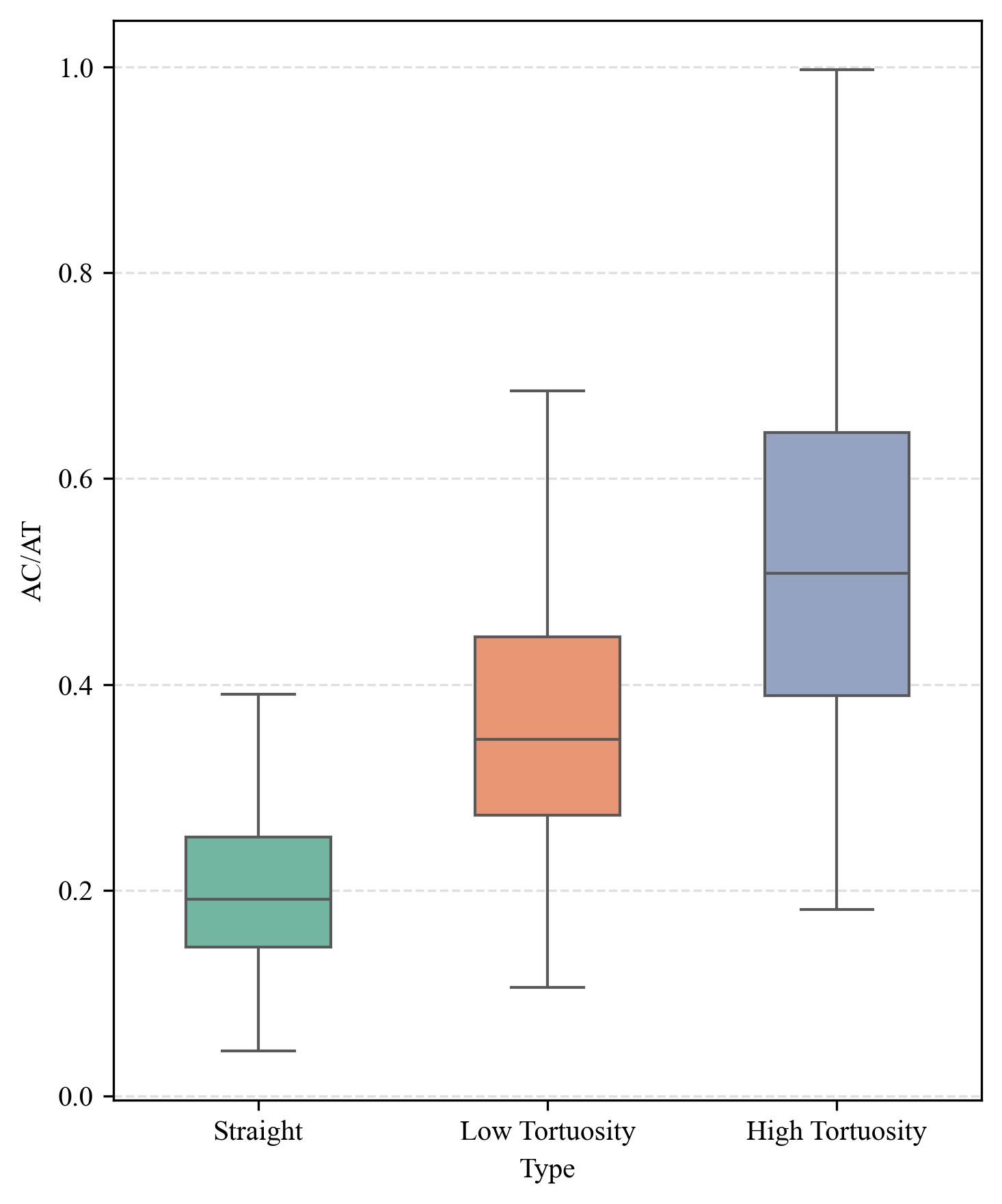} 
        \caption{$\mathcal{AC}/\mathcal{AT}$}
        \label{fig:acat_box}
    \end{subfigure}
    \hfill
    \begin{subfigure}{0.26\textwidth}
        \centering
        \includegraphics[width=\linewidth]{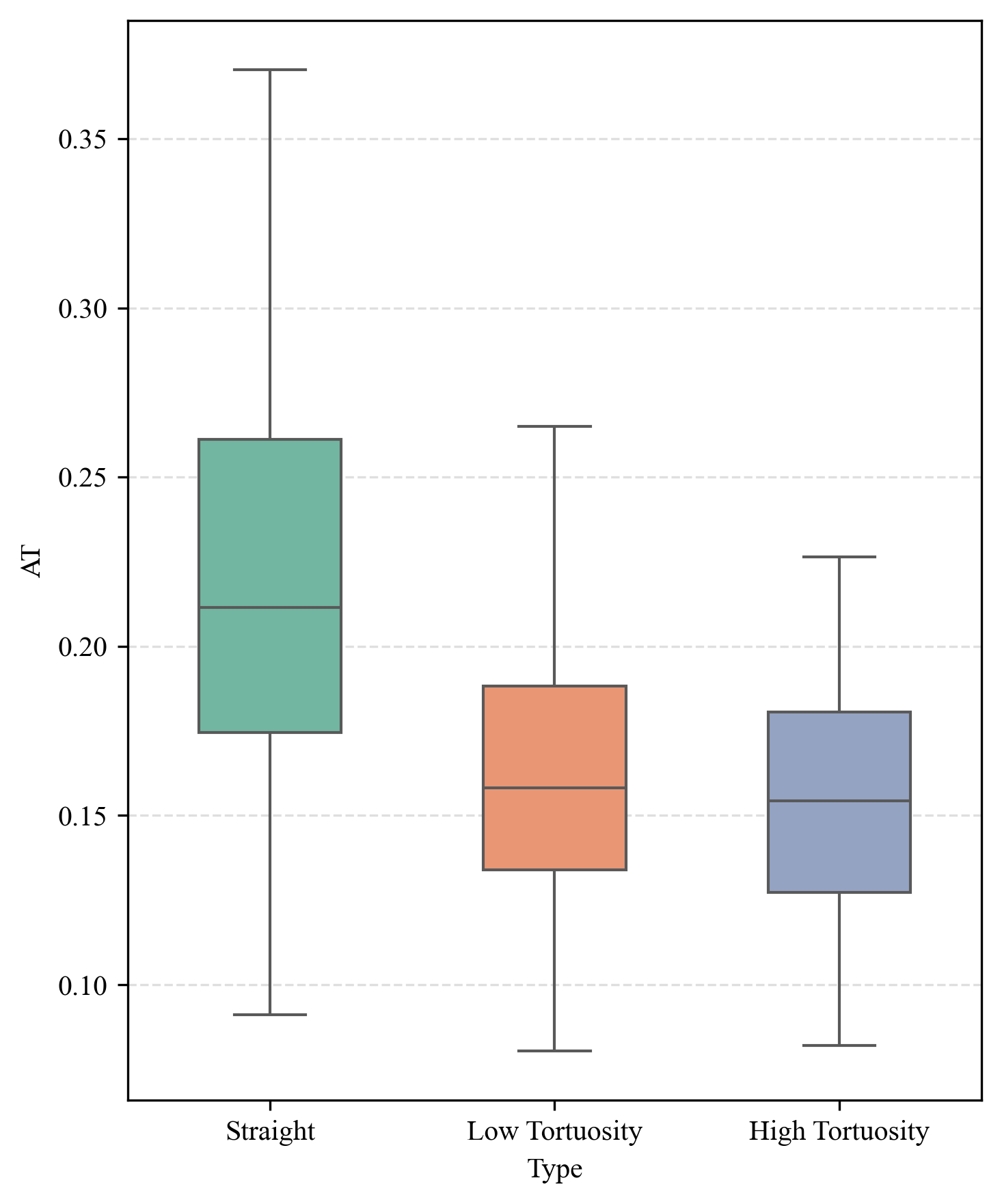} 
        \caption{$\mathcal{AT}$}
        \label{fig:at_box}
    \end{subfigure}
    \hfill
    \begin{subfigure}{0.26\textwidth}
        \centering
        \includegraphics[width=\linewidth]{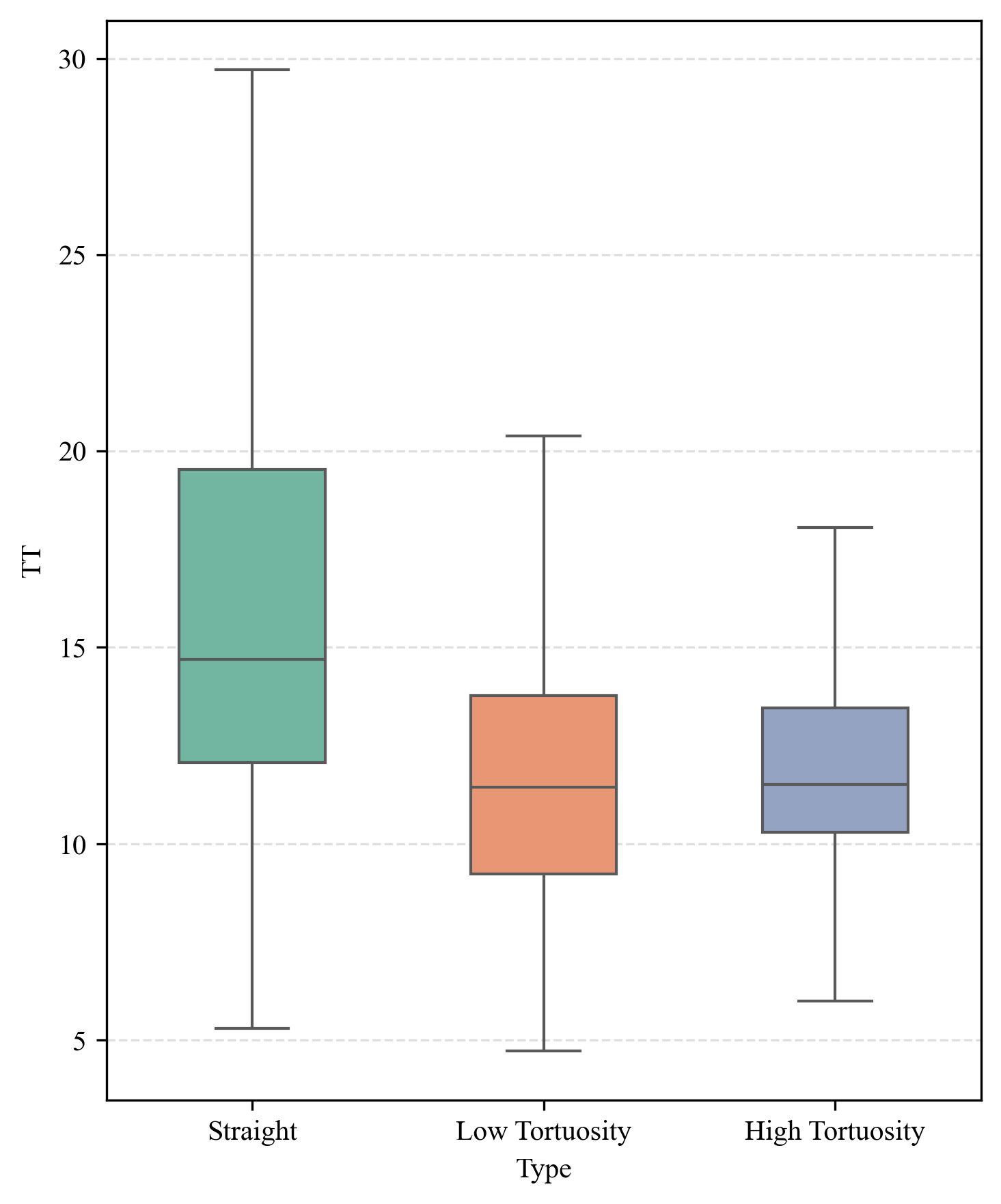} 
        \caption{$\mathcal{TT}$}
        \label{fig:tt_box}
    \end{subfigure}
    
    \caption{Boxplots of the six retained features across the three classes. $\mathcal{TI}$, $\mathcal{AC}$, $\mathcal{TC}$, and $\mathcal{AC}/\mathcal{AT}$ increase across the groups, while $\mathcal{AT}$ and $\mathcal{TT}$ decrease. Extreme points are hidden from the plots to keep the axes readable, but all observations are included in the statistical tests.}
    \label{fig:boxplots_comparison_triple}
\end{figure*}

To substantiate the mathematical basis of clinical assessments, we evaluated the 6 retained features across the three morphological phenotypes (straight, low-tortuosity, and high-tortuosity). As shown in Table~\ref{tab:statistical_comparison}, all six features show significant overall differences among the groups ($p < 0.001$, Kruskal--Wallis H test). The Kruskal--Wallis effect size $\varepsilon^2$ is also reported \cite{Tomczak2014Need}. The macroscopic metric $\mathcal{TI}$ shows the strongest effect ($\varepsilon^2 = 0.7160$) and consistently increases across the three morphological groups. As seen in Fig.~\ref{fig:boxplots_comparison_triple}(a), $\mathcal{TI}$ has extreme upper outliers in the high-tortuosity group, consistent with pronounced spatial elongation.

The microscopic features further reveal how the vessels deform geometrically. The bending features ($\mathcal{AC}$ and $\mathcal{TC}$) increase across the three phenotypes. The average absolute torsion $\mathcal{AT}$ decreases from the straight group to the two tortuosity groups. Total absolute torsion $\mathcal{TT}$ is also lower in both tortuosity groups than in the straight group, although it shows a small increase from the low-tortuosity group to the high-tortuosity group. The interquartile ranges of both torsion measures become narrower. These findings suggest that the high-tortuosity group is characterized by increased bending accompanied by lower measured torsion than the straight group. The biomechanical origin of this pattern requires further investigation. The coupling ratio $\mathcal{AC}/\mathcal{AT}$ also rises sharply ($\varepsilon^2 = 0.5121$), indicating a shift toward bending-dominant geometry in the high-tortuosity group.

In summary, this statistical analysis shows that the visual categories are associated with measurable changes in 3D blood vessel geometry. The 6 features describe differences ranging from macroscopic lengthening to lower measured torsion. These results support their use as inputs for the machine learning classification in the next section.

\subsection{Classification Performance Analysis}
To evaluate the classification performance of the six selected geometric features, four algorithms (RF, LR, SVM, and XGBoost) were tested using the 5-fold nested out-of-fold (OOF) validation strategy.

\subsubsection{Binary Classification Task: }
The primary objective of the binary task is to establish a decision boundary that accurately isolates ``Severe'' vascular tortuosity from ``Non-Severe'' cases. The overarching performance metrics are summarized in Table~\ref{tab:binary_metrics}.

\begin{table}[htbp]
    \centering
    \caption{Performance of Machine Learning Models in Binary Risk Screening}
    \label{tab:binary_metrics}
    \renewcommand{\arraystretch}{1.3} 
    \setlength{\tabcolsep}{12pt}
    \begin{tabular}{l c c c}
    \toprule
    \textbf{Model} & \textbf{Macro-P} & \textbf{Macro-R} & \textbf{Macro-F1} \\
    \midrule
    \textbf{RF}     & \textbf{0.9217} & \textbf{0.9195} & \textbf{0.9206} \\
    XGBoost         & 0.9257 & 0.9143 & 0.9197 \\
    LR              & 0.8909 & 0.9168 & 0.9020 \\
    SVM             & 0.8883 & 0.9123 & 0.8987 \\
    \bottomrule
    \end{tabular}
\end{table}

As illustrated, the tree-based ensemble models (RF and XGBoost) achieved higher Macro-F1 scores than LR and SVM. Specifically, RF achieved the optimal balance with a peak Macro-F1 score of 0.9206. 

Because overall accuracy can obscure minority-class errors in imbalanced classification \cite{Haixiang2017Learning}, we additionally examined severe-case false negatives. Analysis of the consolidated confusion matrices (Fig.~\ref{fig:binary_confusion_matrices}) shows that both RF and XGBoost yielded 354 correct predictions. RF missed 13 severe cases and produced 12 false positives, whereas XGBoost missed 15 severe cases and produced 10 false positives. Thus, RF reduced the false-negative count by two at the cost of two additional false positives.

\begin{figure*}[htbp]
    \centering
    \captionsetup{type=table} 
    \captionof{table}{Consolidated Confusion Matrices in Binary Risk Screening ($N=379$)}
    \label{fig:binary_confusion_matrices}
    \renewcommand{\arraystretch}{1.3}
    \setlength{\tabcolsep}{10pt}
    
  \begin{subtable}{0.48\linewidth}
    \centering
    \caption{RF}
    \label{subtab:bin_conf_rf}
    \begin{tabular}{l c c}
    \toprule
    & Pred Non-Severe & Pred Severe \\
    \midrule
    True Non-Severe & \textbf{255} & 12 \\
    True Severe     & 13 & \textbf{99} \\
    \bottomrule
    \end{tabular}
  \end{subtable}
  \hfill
  \begin{subtable}{0.48\linewidth}
    \centering
    \caption{LR}
    \label{subtab:bin_conf_lr}
    \begin{tabular}{l c c}
    \toprule
    & Pred Non-Severe & Pred Severe \\
    \midrule
    True Non-Severe & \textbf{244} & 23 \\
    True Severe     & 9 & \textbf{103} \\
    \bottomrule
    \end{tabular}
  \end{subtable} 

  \vspace{0.4cm}

  \begin{subtable}{0.48\linewidth}
    \centering
    \caption{SVM}
    \label{subtab:bin_conf_svm}
    \begin{tabular}{l c c}
    \toprule
    & Pred Non-Severe & Pred Severe \\
    \midrule
    True Non-Severe & \textbf{244} & 23 \\
    True Severe     & 10 & \textbf{102} \\
    \bottomrule
    \end{tabular}
  \end{subtable}
  \hfill
  \begin{subtable}{0.48\linewidth}
    \centering
    \caption{XGBoost}
    \label{subtab:bin_conf_xgboost}
    \begin{tabular}{l c c}
    \toprule
    & Pred Non-Severe & Pred Severe \\
    \midrule
    True Non-Severe & \textbf{257} & 10 \\
    True Severe     & 15 & \textbf{97} \\
    \bottomrule
    \end{tabular}
  \end{subtable}

  \vspace{0.2cm}
  
  \begin{flushleft}
    \footnotesize
    \textit{Note:} Rows denote ground-truth morphological statuses, and columns denote predictions. Bold diagonal values indicate correct classifications. LR and SVM each misclassified 23 non-severe cases as severe.
  \end{flushleft}
\end{figure*}

\subsubsection{Ternary Classification Task}
Transitioning from binary classification to detailed ternary grading introduces greater overlap among the three classes. The confusion matrices show that some blood vessels with different labels have similar values for the selected features, which makes the classes harder to separate. The lower Macro-F1 scores across all algorithms reflect this added classification difficulty (Table~\ref{tab:perf_metrics}).

\begin{table}[htbp]
    \centering
    \caption{Quantitative Performance in Ternary Morphological Classification}
    \label{tab:perf_metrics}
    \renewcommand{\arraystretch}{1.3} 
    \setlength{\tabcolsep}{12pt}
    \begin{tabular}{l c c c}
    \toprule
    \textbf{Model} & \textbf{Macro-P} & \textbf{Macro-R} & \textbf{Macro-F1} \\
    \midrule
    \textbf{RF}     & \textbf{0.8637} & \textbf{0.8616} & \textbf{0.8626} \\
    XGBoost         & 0.8573 & 0.8502 & 0.8535 \\
    LR              & 0.8337 & 0.8568 & 0.8402 \\
    SVM             & 0.8333 & 0.8494 & 0.8374 \\
    \bottomrule
    \end{tabular}
\end{table}

RF has been shown to outperform logistic regression across many benchmark datasets \cite{Couronne2018Random}. In the present study, RF achieved the highest Macro-F1 score among the evaluated models (0.8626), exceeding XGBoost by 0.0091 Macro-F1 points. With the matrices arranged as straight, low tortuosity, and high tortuosity, LR and SVM each misclassified 31 high-tortuosity cases as straight, whereas RF made 15 such errors (Table~\ref{tab:sub_confusion_matrices}). Based on these observed results, RF was selected as the core classification model.

\begin{figure*}[htbp]
    \centering
    \captionsetup{type=table} 
    \captionof{table}{Consolidated Confusion Matrices in Ternary Morphological Classification ($N=379$)}
    \label{tab:sub_confusion_matrices}
    \renewcommand{\arraystretch}{1.3}
    \setlength{\tabcolsep}{6pt}
    
  \begin{subtable}{0.48\linewidth}
    \centering
    \caption{RF}
    \label{subtab:conf_rf}
    \begin{tabular}{l c c c}
    \toprule
    & Pred Straight & Pred Low & Pred High \\
    \midrule
    True Straight & \textbf{78} & 2 & 13 \\
    True Low     & 0 & \textbf{99} & 13 \\
    True High    & 15 & 9 & \textbf{150} \\
    \bottomrule
    \end{tabular}
\end{subtable}
\hfill
\begin{subtable}{0.48\linewidth}
    \centering
    \caption{LR}
    \label{subtab:conf_lr}
    \begin{tabular}{l c c c}
    \toprule
    & Pred Straight & Pred Low & Pred High \\
    \midrule
    True Straight & \textbf{86} & 0 & 7 \\
    True Low     & 0 & \textbf{100} & 12 \\
    True High    & 31 & 12 & \textbf{131} \\
    \bottomrule
    \end{tabular}
\end{subtable}

\vspace{0.4cm}

\begin{subtable}{0.48\linewidth}
    \centering
    \caption{SVM}
    \label{subtab:conf_svm}
    \begin{tabular}{l c c c}
    \toprule
    & Pred Straight & Pred Low & Pred High \\
    \midrule
    True Straight & \textbf{84} & 0 & 9 \\
    True Low     & 0 & \textbf{98} & 14 \\
    True High    & 31 & 9 & \textbf{134} \\
    \bottomrule
    \end{tabular}
\end{subtable}
\hfill
\begin{subtable}{0.48\linewidth}
    \centering
    \caption{XGBoost}
    \label{subtab:conf_xgboost}
    \begin{tabular}{l c c c}
    \toprule
    & Pred Straight & Pred Low & Pred High \\
    \midrule
    True Straight & \textbf{74} & 2 & 17 \\
    True Low     & 0 & \textbf{100} & 12 \\
    True High    & 14 & 10 & \textbf{150} \\
    \bottomrule
    \end{tabular}
\end{subtable}
    \vspace{0.2cm}
    \begin{flushleft}
    \footnotesize
    \textit{Note:} Rows and columns denote true and predicted morphological categories, respectively. RF reduces several class-boundary errors compared with LR and SVM. The ternary SVM uses an RBF kernel and is not a linear model.
    \end{flushleft}
\end{figure*}

\subsubsection{Dual-Task Feature Ablation Analysis}

We conducted a backward ablation study to evaluate the structural necessity of the tortuosity features. The results are quantified in Table~\ref{tab:ablation}.

\begin{table*}[htbp]
\centering
\renewcommand{\arraystretch}{1.3} 
\setlength{\heavyrulewidth}{1.5pt} 

\caption{\textbf{[Key Results]} Iterative Feature Elimination and Performance Variation via Backward Ablation}
\label{tab:ablation}

\begin{tabular*}{\textwidth}{@{\extracolsep{\fill}} c l c c @{}}
\toprule
\textbf{Feature Set Size} & \textbf{Retained Features} & \textbf{Binary Macro-F1} & \textbf{Ternary Macro-F1} \\
\midrule
6 (Baseline) & $\mathcal{TI}$, $\mathcal{AC}$, $\mathcal{TC}$, $\mathcal{AC}/\mathcal{AT}$, $\mathcal{AT}$, $\mathcal{TT}$ & 0.9206 & 0.8626 \\
5 ($-\mathcal{TT}$) & $\mathcal{TI}$, $\mathcal{AC}$, $\mathcal{TC}$, $\mathcal{AC}/\mathcal{AT}$, $\mathcal{AT}$ & 0.9223 & \textbf{0.8698} \\
4 ($-\mathcal{AT}$) & $\mathcal{TI}$, $\mathcal{AC}$, $\mathcal{TC}$, $\mathcal{AC}/\mathcal{AT}$ & 0.9172 & 0.8492 \\
3 ($-\mathcal{AC}/\mathcal{AT}$) & $\mathcal{TI}$, $\mathcal{AC}$, $\mathcal{TC}$ & \textbf{0.9262} & 0.8637 \\
2 ($-\mathcal{TC}$) & $\mathcal{TI}$, $\mathcal{AC}$ & 0.8970 & 0.8495 \\
1 ($-\mathcal{AC}$) & $\mathcal{TI}$ & 0.8150 & 0.8185 \\
\bottomrule
\end{tabular*}
\end{table*}

Based on Table~\ref{tab:ablation}, we highlight three key findings regarding feature structural necessity:

\begin{itemize}
\item \textbf{Sufficiency of the Unified 6-Feature Baseline:} The 6 pre-selected features provide a combined description of global elongation, local bending, and spatial twisting. The ablation results show that the binary task peaks at 3 features (0.9262), and the ternary task peaks at 5 features (0.8698). However, the performance gaps between these peaks and the 6-feature baseline (0.9206 and 0.8626) are very small (less than 0.01). Therefore, we choose to retain all 6 features. This unified strategy preserves the full set of geometric descriptors while keeping the performance differences below 0.01.

\item \textbf{Curvature for Binary vs. Torsion for Ternary:} The two tasks benefit from different geometric information. For the binary task, the Macro-F1 score reaches its peak (0.9262) using 3 features ($\mathcal{TI}$, $\mathcal{AC}$, $\mathcal{TC}$), indicating that macroscopic bending features provide strong information for binary detection. In contrast, the ternary task peaks at 5 features (0.8698), which include torsion-related metrics such as $\mathcal{AC}/\mathcal{AT}$. The observed fluctuations after removing torsion-related features suggest that these features provide complementary information for fine-grained ternary subtyping.

\item \textbf{Performance Enhancement Beyond $\mathcal{TI}$ Alone:} Using $\mathcal{TI}$ alone yielded Macro-F1 scores of 0.8150 and 0.8185 for the binary and ternary tasks, respectively. Adding curvature- and torsion-related descriptors improved performance in both tasks, indicating that global elongation alone does not fully represent the geometric information required for vascular morphology classification. These results support the combined use of global elongation, local bending, and non-planar twisting features.

\end{itemize}

\subsubsection{Clinical Translation: Morphological Risk Index (MRI)}

By extracting the impurity-based feature importance values from the Random Forest model, we define two exploratory Morphological Risk Indices (MRI). In this paper, ``risk'' refers only to the model-based blood vessel shape category and does not refer to surgical risk, treatment risk, or patient outcome.

We use the feature importances from the Random Forest model as linear weights. Combining these weights with the standardized features gives the reference formulas for the binary and ternary systems:

\begin{equation}
    \resizebox{\linewidth}{!}{$
    \begin{aligned}
    \text{MRI}_{\mathit{binary}} =& \; 0.3434 \cdot \mathit{AC}_{\mathit{std}} + 0.2411 \cdot \mathit{TI}_{\mathit{std}} + 0.2312 \cdot \mathit{TC}_{\mathit{std}} \\
    &+ 0.0951 \cdot (\mathit{AC}/\mathit{AT})_{\mathit{std}} + 0.0454 \cdot \mathit{TT}_{\mathit{std}} + 0.0438 \cdot \mathit{AT}_{\mathit{std}}
    \end{aligned}
    $}
    \label{eq:mri_bin}
\end{equation}
    
\begin{equation}
    \resizebox{\linewidth}{!}{$
    \begin{aligned}
    \text{MRI}_{\mathit{ternary}} =& \; 0.3273 \cdot \mathit{TI}_{\mathit{std}} + 0.2434 \cdot \mathit{AC}_{\mathit{std}} + 0.1767 \cdot \mathit{TC}_{\mathit{std}} \\
    &+ 0.1275 \cdot (\mathit{AC}/\mathit{AT})_{\mathit{std}} + 0.0644 \cdot \mathit{AT}_{\mathit{std}} + 0.0607 \cdot \mathit{TT}_{\mathit{std}}
    \end{aligned}
    $}
    \label{eq:mri_ter}
\end{equation}

Here, each vessel tortuosity feature in Eqs.~(\ref{eq:mri_bin}) and~(\ref{eq:mri_ter}) must be standardized to remove differences in scale. Each raw feature $x$ is transformed using Z--score normalization:

\begin{equation}
x_{\mathrm{std}} = \frac{x - \mu_x}{\sigma_x}
\label{eq:z_score}
\end{equation}
where $x$ represents a raw blood vessel tortuosity feature of a patient, such as $\mathcal{AC}$, $\mathcal{TI}$, or $\mathcal{TC}$. The parameter $\mu_x$ is the mean value, and $\sigma_x$ is the standard deviation of feature $x$ calculated from all sample data. For external use, fixed values estimated from the training data should be used.

Based on the derived MRI formulas, we summarize the key findings regarding feature distribution, mathematical rationale, and clinical reference ranges:

\begin{itemize}[leftmargin=*]
    \item[i)] \textbf{Exploratory Morphological Reference:} The MRI scores provide sample-based numerical references for the two classification tasks.

    \begin{equation}
    \text{Type}_{\mathit{binary}} = 
    \begin{cases} 
    \text{Non-Severe Tortuosity}, & \text{MRI}_{\mathit{binary}} < \theta_0 \\ 
    \text{Severe Tortuosity}, & \text{MRI}_{\mathit{binary}} \geq \theta_0 
    \end{cases}
    \label{eq:boundary_bin}
    \end{equation}

    Here, the mean MRI scores are $-0.4003$ for the non-severe group and $0.9542$ for the severe group, yielding $\theta_0 = (-0.4003 + 0.9542) / 2 = 0.2770$.
    \begin{equation}
    \text{Type}_{\mathit{ternary}} =
    \begin{cases}
    \text{Straight}, & \text{MRI}_{\mathit{ternary}} < \theta_1 \\
    \text{Low Tortuosity}, & \theta_1 \leq \text{MRI}_{\mathit{ternary}} < \theta_2 \\
    \text{High Tortuosity}, & \text{MRI}_{\mathit{ternary}} \geq \theta_2
    \end{cases}
    \label{eq:boundary_ter}
    \end{equation}

    Here, the mean MRI scores are $-0.7268$ for the straight group, $-0.1841$ for the low-tortuosity group, and $0.8895$ for the high-tortuosity group. The thresholds are $\theta_1 = (-0.7268 + (-0.1841))/2 = -0.4555$ and $\theta_2 = (-0.1841 + 0.8895)/2 = 0.3527$.

    \item[ii)] \textbf{Weight Difference between $\mathcal{TI}$ and $\mathcal{AC}$:} $\mathcal{TI}$ ranks first in univariate Information Gain (IG), but $\mathcal{AC}$ receives a higher impurity-based importance in the MRI formulas. IG measures marginal information, whereas Random Forest importance aggregates reductions in node impurity across the fitted trees. Because $\mathcal{TI}$ and $\mathcal{AC}$ share geometric information, their RF importance values after the other features are included need not follow the IG ranking calculated for each feature separately. Recent work has also shown that surrogate and correlated variables can affect impurity-based Random Forest importance \cite{Voges2023Surrogate}. In the present model, the larger weight assigned to $\mathcal{AC}$ suggests that microscopic curvature contributes complementary information after the other features are considered. This interpretation is model-derived and requires external validation.
    
    \item[iii)] \textbf{Structure of the MRI Formulas:} The derived MRI formulas differ from unweighted linear combinations because their coefficients are obtained from the fitted Random Forest importance values. This construction summarizes the relative importance magnitude assigned to each feature by the model; it does not show the positive or negative direction of the feature effect. However, impurity-based importance can be affected by feature correlation, surrogate variables, and model structure \cite{Voges2023Surrogate}; therefore, the resulting MRI should be regarded as an exploratory model-derived score rather than a validated clinical diagnostic index.
    
    \item[iv)] \textbf{Task-Specific Contribution of Torsion:} The feature weights differ between the two tasks. For the binary task, macroscopic bending features ($\mathcal{AC}$, $\mathcal{TI}$, $\mathcal{TC}$) account for $81.6\%$ of the total weight. For the ternary task, the combined weight of torsion-related features ($\mathcal{AC}/\mathcal{AT}$, $\mathcal{AT}$, $\mathcal{TT}$) increases to $25.3\%$, with $\mathcal{AC}/\mathcal{AT}$ contributing $12.75\%$. These model-derived weights suggest that torsion-related information plays a larger role in the ternary task, but they do not by themselves establish clinical necessity.
\end{itemize}

\section{Discussion}
\label{sec:Discussion}

\subsection{Geometric Complementarity and Feature Optimization}
Our results show that global metrics like the Tortuosity Index ($\mathcal{TI}$) are not enough to describe complex vascular shapes. While $\mathcal{TI}$ measures overall vessel elongation, it cannot tell apart vessels that look similar globally but have different local deformations. The 6-variable feature set solves this by combining global elongation ($\mathcal{TI}$) with localized bending ($\mathcal{AC}$, $\mathcal{TC}$) and spatial twisting ($\mathcal{AT}$, $\mathcal{TT}$) \cite{Piccinelli2009, zhang2021application}.

The feature ablation study (Table~\ref{tab:ablation}) further evaluates this complementarity. For binary screening, a simpler 3-feature subset ($\mathcal{TI}$, $\mathcal{AC}$, $\mathcal{TC}$) reached peak performance (Macro-F1=0.9262). For ternary grading, the model peaked at 5 features (Macro-F1=0.8698). The performance changes after removing torsion-related features suggest that local torsion information contributes to the separation of complex vessel shapes. We decided to keep all 6 features for both tasks. Although the unified 6-feature set yields slightly different scores (0.9206 for binary, 0.8626 for ternary), the performance gaps from the peaks are less than 0.01. Using one unified feature set keeps the mathematical framework consistent and preserves the full set of geometric descriptors.

\subsection{Model Performance and Clinical Interpretation}
Random Forest has shown strong predictive performance relative to logistic regression across many benchmark datasets \cite{Couronne2018Random}. In the present study, RF performed better than LR and the RBF-SVM baseline, suggesting that nonlinear decision boundaries are useful for separating the overlapping geometric patterns in this dataset.

However, nonlinear machine learning models are often regarded as ``black boxes,'' which can hinder clinical acceptance \cite{amann2020explainability,Ghassemi2021FalseHope}. Clinicians require explanations that are actionable, contextualized to clinical workflows, and evaluated with intended users rather than assumed to be transparent solely because a numerical importance score is available \cite{tonekaboni2019what,Chen2022HumanCentered}. To provide a descriptive summary of the RF model, we used its impurity-based feature importance values to create an exploratory Morphological Risk Index (MRI). This uses the fitted model importance values to form explicit, computable scores that can be examined against the underlying geometric features.

The MRI maps the model's impurity-based feature importance values into two quantitative scores (Eqs.~\ref{eq:mri_bin} and \ref{eq:mri_ter}). The feature weights show a shift between tasks: macroscopic bending features account for $81.6\%$ of the binary score and $74.7\%$ of the ternary score, while the combined weight of torsion-related features increases from $18.4\%$ to $25.3\%$. These model-derived distributions suggest that torsion-related information contributes more strongly to the ternary grading task. External validation is required before the MRI thresholds can be interpreted as clinical decision criteria.

\subsection{Clinical Implications and Limitations}
This study provides a quantitative, geometry-based method for evaluating blood vessel shapes. Unlike visual assessment alone, the method provides numerical measurements that can support morphological classification. It does not replace clinical judgment.

However, our study has some limitations. First, this is a retrospective study with a small number of patients from a single center. Small samples and inadequate validation can lead to overly optimistic machine-learning performance estimates \cite{Vabalas2019Machine,Roberts2021Pitfalls}. Because of this, future studies using data from multiple hospitals are needed to confirm that the model works reliably across different imaging machines and scanning protocols. Such studies should follow current guidance for transparent reporting and prospective clinical evaluation of AI-based prediction models \cite{Vasey2022DECIDEAI,Collins2024TRIPODAI,Lekadir2025FUTUREAI}.

Second, the exact clinical meaning of the ``low-tortuosity'' group is not fully clear yet. Future research should track actual surgical outcomes to see if these specific geometric shapes directly predict surgery difficulty or patient risks.

\section{Conclusion}
\label{sec:Conclusion}

This study developed a mathematical framework for the morphological classification of the ICA-C1 segment. The framework combines discrete geometric measurement of blood vessel tortuosity features, Information Gain-based feature selection, and Random Forest classification. It provides a quantitative link between the geometric measurement of 3D blood vessel centerlines and visually defined morphological categories.

The main findings are summarized as follows:
\begin{enumerate}

    \item \textbf{Geometric Description of Blood Vessel Morphology:} Information Gain and correlation analysis identified six geometric features: $\mathcal{TI}$, $\mathcal{AC}$, $\mathcal{TC}$, $\mathcal{AC}/\mathcal{AT}$, $\mathcal{AT}$, and $\mathcal{TT}$. Together, these features describe blood vessel elongation, bending, and twisting while retaining sufficient geometric information for both classification tasks. The statistical results further showed clear differences in these features among the straight, low-tortuosity, and high-tortuosity groups.
    
    \item \textbf{Binary and Ternary Morphological Classification:} Among the evaluated models, RF achieved the highest Macro-F1 score in both tasks. It obtained a Macro-F1 score of 0.9206 for binary classification of non-severe and severe tortuosity and 0.8626 for ternary classification of straight, low-tortuosity, and high-tortuosity blood vessels. The RF feature importance results also showed that elongation- and curvature-related features provided the main information for binary classification. In the ternary task, torsion provides additional information for more detailed morphological classification.
    
    \item \textbf{Morphological Risk Index:} Based on the RF feature importance values, an exploratory Morphological Risk Index was constructed to provide a direct numerical reference for blood vessel morphology. The binary and ternary MRI formulas combine the selected geometric features into one numerical score and provide sample-based reference thresholds for the corresponding morphological categories. Although the MRI does not replace the RF classifier, it offers a simpler way to compare and describe blood vessel morphology quantitatively.

\end{enumerate}

\bibliographystyle{IEEEtranS}
\bibliography{refs}

\end{document}